\documentclass[a4paper,12pt]{article}

\pdfoutput=1 

\usepackage[hmargin=.7in,vmargin=1.1in]{geometry}
\usepackage{indentfirst}
\usepackage{amsfonts}
\usepackage{mathrsfs}
\usepackage{amsmath}
\usepackage{amssymb}
\usepackage{authblk}
\usepackage{cite}
\usepackage{xcolor}
\usepackage{mathtools}
\usepackage{tensor}
\usepackage{physics}
\usepackage{graphicx}
\usepackage{bm}
\usepackage{upgreek}
\usepackage{braket}
\usepackage{color,soul}
\usepackage{csquotes}
\usepackage{caption}
\usepackage{subcaption}

\usepackage[bookmarksnumbered=true,bookmarksopen=true]{hyperref}
 \hypersetup{colorlinks,
             linkcolor=[rgb]{0,0.3,0.6}, 
             citecolor=[rgb]{0,0.3,0.6}, 
             urlcolor=[rgb]{0,0.3,0.6}}

\newcommand\mi{\mathrm{i}}
\newcommand\me{\mathrm{e}}

\newcommand\pp{\uppi}
\newcommand{\dif}{\mathrm{d}}
\newcommand{\ff}{\mathcal{F}}
\newcommand{\ml}{\mathcal{L}}
\DeclareMathOperator{\diag}{diag}

\newcommand\ii{\mathcal{I}}

\begin{document}

\title{\Large\textbf{Regular black holes: A short topic review}}

\author[a,b]{Chen Lan\thanks{stlanchen@yandex.ru}}

\author[b]{Hao Yang\thanks{hyang@mail.nankai.edu.cn}}

\author[b]{Yang Guo\thanks{guoy@mail.nankai.edu.cn}}

\author[b]{Yan-Gang Miao\thanks{Corresponding author: miaoyg@nankai.edu.cn.}}

\affil[a]{\normalsize{\em Department of Physics, Yantai University, 30 Qingquan Road, Yantai 264005, China}}

\affil[b]{\normalsize{\em School of Physics, Nankai University, 94 Weijin Road, Tianjin 300071, China}}

\date{ }

\maketitle

\begin{abstract}
The essential singularity in Einstein's gravity can be avoidable 
if the preconditions of Penrose's theorem can be bypassed, 
i.e., if the strong energy condition is broken in the vicinity of a black hole center. 
The singularity mentioned here includes two aspects: 
(i) the divergence of curvature invariants,
and (ii) the incompleteness of geodesics.
Both aspects are now taken into account in order to determine whether a black hole contains essential singularities.
In this sense, black holes without essential singularities are dubbed regular (non-singular) black holes.
The regular black holes have some intriguing phenomena that are different from those of singular black holes, and such phenomena have inspired numerous studies.
In this review, we summarize the current topics that are associated with regular black holes.
\end{abstract}


\tableofcontents

\section{Introduction}
\label{sec:intr}

Regular black holes (RBHs) are a collection of black holes (BHs) that have coordinate singularities (horizons) 
but lack essential singularities in the entire spacetime.
In most cases, the strategy to determine a RBH refers~\cite{Dymnikova:1992ux,Ayon-Beato:1998hmi,Bronnikov:2000vy} to the spacetime with {\em finite curvature invariants}%
\footnote{The curvature invariants are a set of 
independent scalars that are constructed by a Riemann tensor and a metric \cite{Weinberg:1972kfs},
for instance, the Ricci curvature $R=g^{\mu\nu}R_{\mu\nu}$, the contraction of two Ricci tensors $R_{\mu\nu}R^{\mu\nu}$, and the Kretschmann scalar $R_{\mu\nu\alpha\beta}R^{\mu\nu\alpha\beta}$.} 
everywhere, particularly at the BH center.
This is related to Markov's limiting curvature conjecture \cite{Markov:1982ld,Frolov:1988vj,Frolov:2016pav,Chamseddine:2016ktu}, which states that the curvature invariants must be uniformly restricted by a certain universal value.
However, such a strategy fails~\cite{Misner:1969sf,Kagramanova:2010bk} in the well-known Taub-NUT BH, %
see the end of Sec.\ \ref{sec:curvatures}.
because the null and timelike geodesics are incomplete at the horizon,
which contradicts~\cite{Hawking:1973uf,Wald:1984rg} the alternative strategy to determine a regular spacetime based on the {\em geodesic completeness}.\footnote{According to this strategy, a spacetime is regular if its null and timelike geodesics are complete, i.e., the affine parameter of a test particle will not terminate at any finite value \cite{Hawking:1973uf,Wald:1984rg}.}
The strategy of complete geodesics also encounters~\cite{Carballo-Rubio:2019fnb,Carballo-Rubio:2021wjq} counterexamples, see e.g.\ Refs.\ \cite{Geroch:1968ut,Olmo:2015bya}, where the geodesics are complete but the curvature invariants are divergent, which consequently contradicts Markov's limiting curvature conjecture. 
In this sense, the two strategies should be complementary to each other in order to judge RBHs.

The studies of RBHs can date back to Sakharov and Gliner's works \cite{Sakharov:1966aja,Gliner:1966},
where they stated that the essential singularities can be avoided if the vacuum is replaced by a vacuum-like medium endowed with a de Sitter metric.
This idea has been developed further by  Dynmnikova, Gurevich, and Starobinsky \cite{Gliner:1975nfc,Gurevich:1975om,Starobinsky:1979ty}, see also the reviews \cite{Silbergleit2017,Ansoldi:2008jw}.
The first model of RBHs was implemented by Bardeen \cite{Bardeen:1968nsg}, 
now called the Bardeen BH, which was constructed by simply replacing the mass of Schwarzschild BHs with a $r$-dependent function. As a result, the essential singularity of the Kretschmann scalar is removed in the Bardeen BH, meanwhile, the core of this BH is of de Sitter, i.e., the Ricci curvature is positive in the vicinity of the BH center. 

After three decades of Bardeen's proposal, Ay\'on-Beato and Garc\'ia provided~\cite{Ayon-Beato:2000mjt} the first interpretation of the Bardeen BH in field theory, i.e., they {\em speculated} a source, a magnetic monopole in the context of nonlinear electrodynamics, which can lead to the Bardeen BH solution from Einstein's field equations.
Recently, 
a large number of RBH models have been given in this way. In particular, 
such an approach has been extended~\cite{Bronnikov:2000vy,Fan:2016hvf,Bronnikov:2005gm,Bronnikov:2021uta,Bokulic:2022cyk,Canate:2022zst,Cisterna:2020rkc} to interpret all RBH models with spherical symmetry. It is different from the usual way in finding BH solutions by solving Einstein's field equations. 
According to this approach, one writes the desired RBH and magnetic monopole solutions at first, and then determines the corresponding action of nonlinear electrodynamics. Because of such a special logic, i.e., from the solutions to the action of matters, the RBHs have nontrivial (phantom) scalar hairs \cite{Bronnikov:2005gm,Babichev:2020qpr,Chew:2022enh,Barrientos:2022avi}.
Moreover, they are regarded to be classical objects as they are the solutions of Einstein's field equations.

There are two different ways to construct RBH models: one is to {\em solve} Einstein's field equations that are associated with a kind of special sources, e.g., the matters with spatial distributions \cite{Dymnikova:1992ux,Nicolini:2005vd,Nicolini:2008aj,Spallucci:2008ez,Nicolini:2009gw,Balakin:2006gv,Balakin:2016mnn,Roupas:2022gee};
and the other is to {\em derive} RBHs as quantum corrections to the classical BHs with singularity, e.g., the loop quantum gravity and asymptotic safety method \cite{Bonanno:2000ep,Modesto:2004xx,Gambini:2008dy,Koch:2014cqa,Perez:2017cmj,Bodendorfer:2019jay,Bodendorfer:2019nvy,Bojowald:2020dkb,Brahma:2020eos}. 
Based on the former way, the RBHs behave semiclassically; whereas based on the latter, those RBH models exhibit quantum behaviors. 
In other words, RBHs are now regarded as a tool to study the classical limit of quantum black holes
because we do not have a complete theory of quantum gravity at the moment.

Besides the structures of RBHs \cite{Borde:1996df,Bronnikov:2006fu,Zaslavskii:2010qz,Carballo-Rubio:2018pmi,Bonanno:2020fgp,Li:2021bmj,Carballo-Rubio:2021bpr,Giacchini:2021pmr},
the study  also extends to the other areas of BH physics, including thermodynamics \cite{Fan:2016hvf,Fan:2016rih,Lan:2020fmn,Bouhmadi-Lopez:2020wve,Guo:2021wcf}, dynamics  \cite{Flachi:2012nv,Cai:2020kue,Li:2021qim,Cai:2021ele,Li:2021epb,Guo:2022hjp}, shadows \cite{Li:2013jra,Abdujabbarov:2016hnw,Tsukamoto:2017fxq,Dymnikova:2019vuz,Kumar:2019pjp,Ghosh:2020ece,Jusufi:2020odz,Guo:2021wid,Ling:2022vrv,KumarWalia:2022aop},
quasinormal modes \cite{Bronnikov:2012ch,Flachi:2012nv,Li:2013fka,Fernando:2012yw,Toshmatov:2015wga,Toshmatov:2017bpx,Toshmatov:2018tyo,Toshmatov:2018ell,Konoplya:2022hll,Lan:2022qbb,Konoplya:2023aph}, superradiance 
\cite{Yang:2022uze},
and synchrotron
radiations \cite{Liu:2022ruc,Riaz:2022rlx,Riaz:2023yde} , etc.
All of the research on RBHs aims to explore how RBHs differ from singular black holes (SBHs) and thus shed light on further research for quantum gravity.
Currently, the study of RBHs has made great progress in depth and breadth, which leads to the necessity for us to summarize the new results in a systematic way.
Comparing with the previous reviews \cite{Ansoldi:2008jw,Nicolini:2008aj,Torres:2022twv}, we would like to gather the new progress developed recently. 
This is the main motivation for drafting this review.

Our review is organized as follows.
In Sec.\ \ref{sec:construction}, we address the issue of the construction for both the non-rotating and rotating RBHs,
where we also discuss the number of curvature invariants among the Zakhary-Mcintosh invariants (a complete set of curvature invariants) for determining whether a BH is regular or not.
Sec.\ \ref{sec:interpretation} includes the clarification for understanding RBHs and the establishment of the sources of RBHs in terms of Petrov's approach. We end this section by showing the peculiarity of RBHs on scalar hairs, which does not break the non-hair theorem.
In Sec.\ \ref{sec:energyconditions} we demonstrate the role played by the strong energy condition in RBHs, and provide an illustration or a resolution of the issue of the violation of the other energy conditions in RBHs. 
Sec.\ \ref{sec:thermodynamics} is dedicated to the thermodynamics of RBHs, where
we give a discussion on the entropy-area law, 
based on which the self-consistent first law of thermodynamics is given in Sec.\ \ref{sec:themlaw}.
In Sec.\ \ref{sec:bhchemandruppgeo}, we discuss  thermodynamic geometry for RBHs because they are the nontrivial extensions of interesting issues associated with SBHs. 
Finally, we conclude in Sec.\ \ref{sec:conclusion} with some outlooks.

\section{Construction of regular black holes}
\label{sec:construction}

In this section, we summarize the approaches for both the non-rotating and rotating RBHs, meanwhile, we analyze the minimum set of curvature invariants that are needed to judge an RBH.
Here the curvature invariants  refer to the {\em Zakhary-Mcintosh (ZM) invariants}  \cite{Zakhary:1997acs,Overduin:2020cif} 
that form a complete set of Riemann invariants and contain seventeen elements, see also App.\ \ref{app:zm-invariants}. 
The reason that we adopt the ZM invariants rather than the usual ones, such as the Ricci scalar and Kretschmann scalar \cite{Balart:2014cga,Fan:2016hvf,Frolov:2016pav}, will be explained in Sec.\ \ref{sec:curvatures}.

\subsection{How to construct non-rotating regular black holes?}
\label{sec:nonrotating}

Despite the complexity of ZM invariants, the calculation of ZM invariants becomes simple for those BHs with spherical symmetry.
The RBHs with spherical symmetry have two types of metrics:
A metric in the first type involves one shape function, see the specific square of line elements,
\begin{equation}
\label{eq:sphe-met-1}
    \dif s^2 = -f(r) \dif t^2
    +f^{-1}(r) \dif r^2 +r^2\dif \Omega^2,
\end{equation}
and a metric in the second type involves two shape functions, see the specific square of line elements,
\begin{equation}
\label{eq:sphe-met-12}
    \dif s^2 = -f(r) \dif t^2
    +f^{-1}(r) A^2(r)\dif r^2 +r^2\dif \Omega^2,
\end{equation}
which is equivalent to 
\begin{equation}
\label{eq:sphe-met-2}
      \dif s^2 = -f(\xi) \dif t^2
    +f^{-1}(\xi) \dif \xi^2 +r^2(\xi)\dif \Omega^2,
\end{equation}
where $\xi$ is a newly defined variable,
\begin{equation}
    \xi\equiv \int \dif r\, A(r). 
\end{equation}
In the following of this subsection,
we shall give some restrictions to the two types of metrics, which will reveal that the RBHs depicted by the two types of metrics have finite curvature invariants.

\subsubsection{The case with one shape function}

It is quite general for us to write the shape function as follows,
\begin{equation}
\label{eq:shape1}
    f(r)=1-\frac{2M \sigma(r)}{r},
\end{equation}
where $\sigma(r)$ is a function of the radial variable $r$ and $M$ is BH mass. 
In order to observe the regularity, we
expand $\sigma(r)$ by the power series around $r=0$,
\begin{equation}
\label{eq:expd-mass}
    \sigma(r)=  \sigma_1 r+ \sigma_2 r^2+ \sigma_3 r^3 +O(r^4),
\end{equation}
where $\sigma_i$ are constant coefficients.
Then, by substituting Eqs.\ \eqref{eq:sphe-met-1}, \eqref{eq:shape1} and \eqref{eq:expd-mass} into the ZM invariants, 
we can find the conditions of finite curvatures, that is, the coefficients $\sigma_1$ and $\sigma_2$ must vanish,
\begin{equation}
    \sigma_1 =0=\sigma_2.
\end{equation}
As an example, we write the behaviors of three usual candidates among the seventeen ZM invariants around $r=0$: The Ricci scalar $R$, the Weyl scalar $W=W^{\alpha\beta\mu\nu}W_{\alpha\beta\mu\nu}$, and the Kretschmann scalar $K$ have the asymptotic behaviors for RBHs,
\begin{equation}
    R=24 M \sigma_3+O(r),\qquad
    W=O(r^2),\qquad
    K=96 M^2 \sigma_3^2+O(r).
\end{equation}

Alternatively, we can select three curvatures from the seventeen ZM invariants and write $\sigma(r)$, $\sigma'(r)$, and $\sigma''(r)$ as functions of these three curvatures because the ZM invariants contain $\sigma(r)$ and only its first and second order derivatives, $\sigma'(r)$ and $\sigma''(r)$. Then, requiring the finiteness of the three curvatures, we can find the behavior of $\sigma(r)$ around the center $r=0$, i.e., $\sigma(r)$ should not decrease slower than $r^3$ as $r$ approaches to zero \cite{Lan:2021ngq}, otherwise, some of the ZM invariants will diverge at $r=0$.

\subsubsection{The case with two shape functions}

For the RBHs with two shape functions \cite{Frolov:2016pav,Bronnikov:2005gm,Simpson:2018tsi,Boos:2021kqe}, 
we apply a similar procedure to the above, that is, we expand both of shape functions (see Eq.~(\ref{eq:sphe-met-12})) by the power series, 
\begin{subequations}
\begin{equation}
    A(r)=A_0 +A_1 r +A_2 r^2 +O(r^3),\label{Aexpand}
\end{equation}
\begin{equation}
    f(r)=B_0 +B_1 r +B_2 r^2 +O(r^3).\label{frexpand}
\end{equation}
\end{subequations}
After substituting Eqs.~(\ref{eq:sphe-met-12}), (\ref{Aexpand}) and (\ref{frexpand}) into the ZM invariants, we can find the conditions for finite curvatures,
\begin{equation}
    A_0=B_0, \qquad
    A_1=B_1=0,
\end{equation}
i.e., the first order of $r$ must be absent in the power expansions.
We still give three curvature invariants for RBHs as an example when $r$ goes to zero,
\begin{equation}
    R=\frac{6(A_2-2B_2)}{A_0}+O(r),\qquad
    W=O(r^2),\qquad
    \mathcal{S}=\frac{3 A_2^2}{A_0^2}+O(r),
\end{equation}
where $\mathcal{S}$ and $\mathcal{S}_{\mu\nu}$ are defined~\cite{Frolov:2016pav} by $\mathcal{S}\equiv \mathcal{S}^{\mu\nu}\mathcal{S}_{\mu\nu}$ and
 $\mathcal{S}_{\mu\nu} \equiv R_{\mu\nu}-g_{\mu\nu}R/4$, respectively.

\subsection{How many curvature invariants do we need to define a regular black hole?} 
\label{sec:curvatures}

Generally, the {\em finite curvature invariants} and  {\em geodesic completeness} are not equivalent to each other,\footnote{For certain cases, these two conditions are equivalent, e.g., for spherically symmetric BHs with one shape function, i.e., $\dif s^2=-f(r)\dif t^2 +f^{-1}(r) \dif r^2+r^2\dif\Omega^2$.} but they can be regarded as two independent necessary conditions for checking whether a BH is regular.
 Moreover, the former is coordinate-independent, i.e., a coordinate singularity does not appear in curvature invariants, whereas in order to prove the regularity by the latter, one has to eliminate the coordinate singularity by selecting an appropriate coordinate. Thus, the latter often involves the choice of a coordinate system in practice.
For instance, in the Rindler spacetime \cite{Wald:1984rg,Carroll:2004st}, $\dif s^2 = -z^2\dif t^2 +\dif x^2+\dif y^2 +\dif z^2$,
the geodesics cannot be extended along the $z$-direction because the corresponding affine parameter is finite at $z=0$. In other words, the point $z=0$ acts as a singularity in this spacetime.
However, after an appropriate transformation, 
\begin{equation}
t\to \tanh^{-1}\frac{T}{Z},\qquad
x\to X,\qquad 
y\to Y,\qquad
z\to \sqrt{Z^2-T^2},
\end{equation}
the original metric converts to that of the Minkowski spacetime, $\dif s^2 = -\dif T^2 +\dif X^2+\dif Y^2 +\dif Z^2$, i.e., there is no singularity anywhere.
From this point of view, we can see that the condition of {\em finite curvature invariants} has its advantage, i.e., it does not need to be concerned about selecting appropriate coordinates.

Nevertheless, 
there are two questions
associated with the criterion of finite curvature invariants. The first is whether the curvature invariants can reveal the singularity of spacetime, and the
second is how many curvature invariants have to be used in order to determine a RBH if the first question has a positive answer.

It is known that the components of Riemann tensors are not suitable to describe spacetime \cite{dInverno:2022gxs}
because they depend on the choices of coordinate systems. However, the scalars constructed by Riemann tensors and metrics can improve the situation. 
These curvature scalars are important for investigating singularities.
In other words, the curvature scalars are believed to describe the primary properties of spacetime, in particular,
 to determine the existence of spacetime singularities.

Considering the independent components of Riemann tensors and metrics together with the constraints from coordinate transformation, one can construct $14$ curvature scalars\footnote{This number comes from $20$ independent components of a Riemann tensor plus $10$ independent components of a metric but minus $16$ constraints imposed by the general coordinate transformation.} in the $4$D spacetime \cite{Weinberg:1972kfs}.
For simple cases, there exist only three curvature scalars\footnote{Alternatively, $K$ and $R_2$ are replaced by  $W\equiv C_{\mu\nu\alpha\beta}C^{\mu\nu\alpha\beta}$, the contraction of two Weyl tensors, and $\mathcal{S}\equiv\mathcal{S}_{\mu\nu}\mathcal{S}^{\mu\nu}$, where $\mathcal{S}_{\mu\nu}\equiv R_{\mu\nu}-g_{\mu\nu} R/4$ \cite{Frolov:2016pav}.} that are connected by the Ricci decomposition, 
i.e., the Ricci scalar ($R$), the Kretschmann scalar ($K$) and the contraction of two Ricci tensors ($R_2$).

For certain cases in the presence of matter, the $14$ scalars are not {\em complete},  
i.e., more than $14$ scalars are required \cite{Carminati:1991ai}.
 Here the {\em completeness} implies the minimal number of invariants for all $6$ Petrov types and 15 Segr\'e types \cite{Stephani:2003tm}.
It has been proved \cite{Zakhary:1997acs} that the complete set of curvature invariants should contain $17$ elements that are known as ZM invariants. 
These invariants are originally defined by the spinorial quantities in the Penrose-Newman formalism, 
and are recast in the explicit algebraic expressions in Ref.\ \cite{Overduin:2020cif}.

Thus, the two questions mentioned above become whether the ZM invariants can determine the singularities of spacetime and how many elements are required in this set.
The second question has been analyzed for several cases, for instance, 
four scalars are necessary for rotating RBHs \cite{Torres:2016pgk}, while two scalars are enough for non-rotating ones \cite{Hu:2023iuw}.
As to the first question, the situation is relatively complicated and no definite answer is given at the moment.

 Let us see the well-known Taub-NUT BH \cite{Stephani:2003tm,Griffiths:2009dfa} as a sample,
\begin{equation}
    \dif s^2
    =-f(r) \left[\dif t+
    2n \cos(\theta) \dif \phi
    \right]^2
    +\frac{\dif r^2 }{f(r)}
    +\zeta ^2 \left[\dif \theta^2+\sin ^2(\theta ) \dif\phi^2 \right]
\end{equation}
with
\begin{equation}
    f(r)=\frac{\Delta}{\zeta^2},\qquad
    \Delta =r^2-2 M r-n^2,
    \qquad
    \zeta =\sqrt{r^2+n^2},
\end{equation}
where $M$ is mass, the NUT parameter $n$ is positive and referred to as magnetic mass, and
the horizon is located at $r_{\rm H} =M+\sqrt{M^2+n^2}$. 
 Now the Taub-NUT BH is regarded~\cite{Huang:2019cja,Emond:2020lwi} as the electric-magnetic duality of Schwarzschild BHs.
It is shown \cite{Kagramanova:2010bk,Misner:1963fr} that the geodesics are incomplete at the horizon.

We turn to the investigation of the curvature invariants of the Taub-NUT BH.
The Ricci tensor vanishes, $R_{\mu\nu}=0$. The Kretschmann scalar reads
\begin{equation}
    K\sim \frac{48 \left(n^2-M^2\right)}{n^6}+O\left(r\right),
\end{equation}
which is finite as $r$ approaches zero. Moreover, $R_2$ is also finite when $r$ goes to zero. As a result, 
the three curvature invariants, $R$, $K$, and $R_2$ are finite in the Taub-NUT BH spacetime. 
As done for the center $r=0$, we can show that the horizon has no singularity when we check $R$, $K$, and $R_2$.
Further, we confirm that all the ZM invariants are regular everywhere in the Taub-NUT BH spacetime, 
i.e., the finite-curvature method cannot ensure the completeness of geodesics~\cite{Misner:1963fr,Kagramanova:2010bk}.
This implies that the ZM curvature invariants are not able to fully reflect the singularity of spacetime, or more invariants than the ZM ones are needed.
The validity of these statements remains open at the moment.

\subsection{How to construct rotating regular black holes?}

It is considerably difficult to obtain rotating RBH's solutions from the Einstein field equations because the complexity of Einstein's field equations in the case of rotation is much greater than that of the static case.
Therefore, the widely-used method for constructing rotating BHs is the Newman-Janis algorithm (NJA)~\cite{Newman:1965}.

The NJA originated from the connection between a static BH and a rotating one in general relativity.
It is well known that the Schwarzschild, Reissner-Nordstr\"om (RN), Kerr, and Kerr-Newman (KN) BHs were obtained by solving Einstein's field equations in electro-vacuum. These solutions have clear physical explanations.
By comparing the metrics of these BHs, Newman and Janis proposed the NJA to mathematically describe the transformation from spherically symmetric Schwarzschild BHs to axially symmetric Kerr BHs. 
The algorithm can also describe the transformation from RN BHs to KN BHs.

Subsequently, G\"urses and G\"ursey extended \cite{Gurses:1975} this algorithm for the Kerr-Schild type of BHs, namely, the metric of one shape function mentioned in Sec.~\ref{sec:nonrotating}.
Further, Drake and Szekeres generalized~\cite{Drake:1998gf} the NJA for general spherically symmetric BHs.
Based on the above algorithms, many spherically symmetric RBHs have been extended to their axially symmetric counterparts, such as the noncommutative BHs \cite{Nicolini:2008aj,Smailagic:2010nv,Modesto:2010rv}, loop quantum corrected BHs  \cite{Modesto:2008im,Caravelli:2010ff}, Bardeen BHs \cite{Bardeen:1968nsg,Bambi:2013ufa}, and Hayward BHs \cite{Hayward:2005gi,Bambi:2013ufa}, phantom RBHs \cite{Kamenshchik:2023bzc}, etc.

\subsubsection{What are the problems faced by the Newman-Janis algorithm?}\label{sec:problemofNJA}

The NJA faces the uncertainty of complex transformation of metrics.
This problem arises from the complex transformation of coordinates \cite{Azreg-Ainou:2014aqa,Azreg-Ainou:2014pra,Azreg-Ainou:2011tkx}:
\begin{equation}\label{eq:complex-trans}
r\rightarrow r+\mi a\cos\theta,\qquad  u\rightarrow u-\mi a\cos\theta,
\end{equation}
where $(u,r,\theta,\varphi)$ are the advanced null coordinates and $a$ is the rotation parameter. 
According to the transformation, we need to convert a static and spherically symmetric metric function into a rotational and axially symmetric function, and ensure that the latter is real but not complex.
As a result, this conversion of metric functions must follow certain rules.
However, such rules are ambiguous in the NJA. 

The commonly-used rule is obtained by comparing the RN metric with the KN metric.
The $tt$-component of the RN metric reads
\begin{equation}
g_{(\rm{RN})tt}=1-\frac{2M}{r}+\frac{q^2}{r^2},
\end{equation}
where $M$ is mass and $q$ is charge.
The $tt$-component of the KN metric takes the form,
\begin{equation}
g_{(\rm{KN})tt}=1-\frac{2Mr}{r^2+a^2\cos^2\theta}+\frac{q^2}{r^2+a^2\cos^2\theta}.
\end{equation}
The conversion rule is as follows:
\begin{equation}\label{eq:conversion rule}
r^2=rr^*\rightarrow (r+ia\cos\theta)(r-ia\cos\theta)=r^2+a\cos^2\theta,
\end{equation}
\begin{equation}
\frac{1}{r}=\frac{1}{2}\left(\frac{1}{r}+\frac{1}{r^*}\right)\rightarrow \frac{1}{2}\left(\frac{1}{r+ia\cos\theta}+\frac{1}{r-ia\cos\theta}\right)=\frac{r}{r^2+a^2\cos^2\theta}.
\end{equation}

However, the above conversion rule may not be applicable to RBH metrics due to their complexity.
Here we take the black-bounce spacetime \cite{Simpson:2018tsi} as an example, where its metric's $tt$-component reads
\begin{equation}
g_{(\rm{BB})tt}=1-\frac{2M}{\sqrt{r^2+l^2}},
\end{equation}
where $l$ is the regularization parameter. 
When $l$ vanishes, the metric becomes the form of Schwarzschild's spacetime.
Therefore, the rotation formulation of this metric should reduce to the Kerr metric when $l=0$.
But this is not the case. The rotating black-bounce metric under the above conversion rule takes the form,
\begin{equation}
g_{(\rm{rBB})tt}=1-\frac{2M}{\sqrt{r^2+a^2\cos^2\theta+l^2}},
\end{equation}
whereas the $tt$-component of the Kerr metric is
\begin{equation}
g_{(\rm{K})tt}=1-\frac{2Mr}{r^2+a^2\cos^2\theta}.
\end{equation}
Obviously, $g_{(\rm{rBB})tt}$ does not reduce to $g_{(\rm{K})tt}$ when $l=0$.
As a result, the above conversion rule does not apply to the black-bounce spacetime.
Owing to the failure of this rule, we need to find such a rule that applies to more models.
Furthermore, the ambiguity related to coordinate transformations leads to difficulties in the generalization of the NJA.

\subsubsection{How to modify the Newman-Janis algorithm?}
 
 In order to avoid the ambiguity caused by the complex transformation, Azreg-Ainou modified~\cite{Azreg-Ainou:2014pra,Azreg-Ainou:2014aqa} the NJA as follows.

For a general static metric,
\begin{equation}
\dif s^2=-G(r)\dif t^2+\frac{\dif r^2}{F(r)}+H(r)\left(\dif\theta^2+\sin^2\theta\dif\varphi^2\right),
\end{equation}
one introduces the advanced null coordinates $(u,r,\theta,\varphi)$ defined by
\begin{equation}
\dif u=\dif t-\frac{\dif r}{\sqrt{FG}},
\end{equation}
and expresses the contravariant form of the metric in terms of a null tetrad,
\begin{equation}
g^{\mu\nu}=-l^\mu n^\nu-l^\nu n^\mu+m^\mu m^{*\nu}+m^\nu m^{*\mu},
\end{equation}
where 
\begin{subequations}\label{eq:tetrad}
\begin{equation}
l^\mu=\delta^\mu_r,
\end{equation}
\begin{equation}
n^\mu=\sqrt{\frac{F}{G}}\delta^\mu_u-\frac{F}{2}\delta^\mu_r,
\end{equation}
\begin{equation}
m^\mu=\frac{1}{\sqrt{2H}}\left(\delta^\mu_\theta+\frac{i}{\sin\theta}\delta^\mu_\varphi\right),
\end{equation}
\begin{equation}
l_\mu l^\mu=m_\mu m^\mu=n^\nu n_\nu=l_\mu m^\mu=n_\mu m^\mu=0,
\end{equation}
\begin{equation}
l_\mu n^\mu=-m_\mu m^{*\mu}=1.
\end{equation}
\end{subequations}
Then, one introduces the rotation via the complex transformation, Eq.~\eqref{eq:complex-trans}, under which $\delta^\mu_\nu$ transform as a vector:
\begin{equation}\label{eq:transform-delta}
\delta^\mu_r\rightarrow \delta^\mu_r,
\qquad \delta^\mu_u\rightarrow\delta^\mu_u,
\qquad \delta^\mu_\theta\rightarrow\delta^\mu_\theta+ia\sin\theta(\delta^\mu_u-\delta^\mu_r),
\qquad \delta^\mu_\varphi\rightarrow\delta^\mu_\varphi.
\end{equation}

For an SBH, the metric function of its rotating counterpart can be determined under the above complex transformation.
However, such a transformation does not work well for a RBH as we discussed in Sec.~\ref{sec:problemofNJA}.
Thus, one assumes that $\{G,F,H\}$ transform to $\{A,B,\Psi\}$:
\begin{equation}\label{eq:transform-function}
\{G(r),F(r),H(r)\}\rightarrow\{A(r,\theta,a),B(r,\theta,a),\Psi(r,\theta,a)\},
\end{equation}
where $\{A,B,\Psi\}$ are real functions to be determined, and they should recover their static counterparts in the limit $a\rightarrow 0$, namely, 
\begin{equation}
\lim_{a\to 0}A(r,\theta,a)=G(r),\qquad \lim_{a\to 0}B(r,\theta,a)=F(r),\qquad \lim_{a\to 0}\Psi(r,\theta,a)=H(r).
\end{equation}
According to Eq.~\eqref{eq:transform-delta} and Eq.~\eqref{eq:transform-function}, the null tetrad becomes
\begin{subequations}
\begin{equation}
l^\mu=\delta^\mu_r,
\end{equation}
\begin{equation}
n^\mu=\sqrt{\frac{B}{A}}\delta^\mu_u-\frac{B}{2}\delta^\mu_r,
\end{equation}
\begin{equation}
m^\mu=\frac{1}{\sqrt{2\Psi}}\left[\delta^\mu_\theta+ia\sin\theta(\delta^\mu_u-\delta^\mu_r)+\frac{i}{\sin\theta}\delta^\mu_\varphi\right],
\end{equation}
\end{subequations}
and the corresponding metric with rotation takes the form,
\begin{equation}
\begin{split}
\dif s^2=& -A\dif u^2-2\sqrt{\frac{A}{B}}\dif u\dif r-2a\sin^2\theta\left(\sqrt{\frac{A}{B}}-A\right)\dif u\dif \varphi+2a\sin^2\theta\sqrt{\frac{A}{B}}\dif r\dif \varphi\\
& +\Psi\dif\theta^2+\sin^2\theta\left[\Psi+a^2\sin^2\theta\left(2\sqrt{\frac{A}{B}}-A\right)\right]\dif\varphi^2.
\end{split}
\end{equation}

Next, one rewrites the above metric with the Boyer-Lindquist coordinates, and lets the metric have only one off-diagonal term $g_{t\phi}$.
To reach the goal, one needs the following coordinate transformation,
\begin{equation}
\dif u=\dif t+\lambda(r)\dif r,\quad \dif\varphi=\dif\phi+\chi(r)\dif r,
\end{equation}
where $\{\lambda(r),\chi(r)\}$ must depend only on $r$ to ensure integrability.
If the transformation Eq.~\eqref{eq:transform-function} is a priori determined, $\{\lambda(r),\chi(r)\}$ may not exist.
Considering these constraints, one has the formulations of $\{A(r,\theta,a), B(r,\theta,a), \lambda(r), \chi(r)\}$,
\begin{subequations}
\begin{equation}
A(r,\theta)=\frac{(FH+a^2\cos^2\theta)\Psi}{(K+a^2\cos^2\theta)^2},
\end{equation}
\begin{equation}
B(r,\theta)=\frac{FH+a^2\cos^2\theta}{\Psi},
\end{equation}
\begin{equation}
\lambda(r)=-\frac{K+a^2}{FH+a^2},
\end{equation}
\begin{equation}
\chi(r)=-\frac{a}{FH+a^2},
\end{equation}
\end{subequations}
where $K(r)$ is defined by
\begin{equation}
K(r)\equiv\sqrt{\frac{F(r)}{G(r)}}H(r).
\end{equation}
As a result, one obtains the metric for rotating RBHs with the Kerr-like form,
\begin{equation}
\dif s^2=\frac{\Psi}{\rho^2}\left[-\left(1-\frac{2f}{\rho^2}\right)\dif t^2+\frac{\rho^2}{\Delta}\dif r^2-\frac{4af\sin^2\theta}{\rho^2}\dif t\dif \phi+\rho^2\dif\theta^2+\frac{\Sigma\sin^2\theta}{\rho^2}\dif\phi^2\right],
\end{equation}
where
\begin{equation}
\begin{split}
&\rho^2\equiv K+a^2\cos^2\theta,\qquad 2f(r)\equiv K-FH\\
&\Delta(r)\equiv FH+a^2,\qquad \Sigma\equiv(K+a^2)^2-a^2\Delta\sin^2\theta.
\end{split}
\end{equation}

In the above metric, $\Psi(r,\theta,a)$ remains unknown and may be determined by some specific physical interpretations.
For example, if the source is interpreted as an imperfect fluid rotating about the $z$ axis, $\Psi$ obeys~\cite{Azreg-Ainou:2014aqa} the Einstein field equations, 
\begin{equation}
\left(K+a^2y^2\right)^2\left(3\Psi_{,r}\Psi_{,y^2}-2\Psi\Psi_{,ry^2}\right)=3a^2K_{,r}\Psi^2,
\end{equation}
\begin{equation}
\left[K_{,r}^2+K(2-K_{,rr})-a^2y^2(2+K_{,rr})\right]\Psi+(K+a^2y^2)(4y^2\Psi_{,y^2}-K_{,r}\Psi_{,r})=0,
\end{equation}
where “,” in the subscript of variables means derivatives and $y\equiv \cos\theta$.
However, it is almost impossible to determine $\Psi(r,\theta,a)$ in this way because of the high complexity.
For the RBH metrics mentioned in Sec.~\ref{sec:nonrotating}, one usually chooses
\begin{equation}
\Psi(r,\theta,a)=H(r)+a^2\cos^2\theta.\label{psichoice}
\end{equation}
While doing so may lose a reasonable physical explanation, it needs to be tested case by case whether such a choice really loses a physical explanation.
It is worth mentioning that Eq.~(\ref{psichoice}) is compatible with the NJA and is available to construct a rotating RBH, but it is still unclear whether Eq.~(\ref{psichoice}) is the only choice or not.

\subsubsection{What are the regularity conditions of rotating regular black holes?}

 For the seed metric with one shape function, the metric of rotating RBHs takes the form via the NJA,
\begin{equation}\label{eq:rmetric-ss}
\dif s^2=-\frac{\Delta}{\rho^2}(\dif t-a\sin^2\theta\dif \phi)^2+\frac{\rho^2}{\Delta}\dif r^2+\rho^2\dif \theta^2+\frac{\sin^2\theta}{\rho^2}
\left[a\dif t-(r^2+a^2)\dif\phi\right]^2,
\end{equation}
where
\begin{equation}
\rho^2=r^2+a^2\cos^2\theta,\qquad \Delta=r^2-2M\sigma(r)r+a^2.
\end{equation}
This metric reduces to the Kerr metric when $\sigma(r)=1$, and to the KN metric when $\sigma(r)=1-q^2/(2Mr)$. 

Further, the metric Eq.~(\ref{eq:rmetric-ss}) belongs~\cite{Torres:2016pgk,Torres:2022twv} to the Petrov type {\bf{D}} because $\Psi_2$ can be the only non-vanishing scalar when $\Psi_0, \Psi_1, \Psi_3$ and $\Psi_4$ vanish simultaneously, where the five complex scalar functions can be expressed by the Weyl tensor $C_{\kappa\lambda\mu\nu}$ as follows:
\begin{subequations}
\begin{equation}
\Psi_0=C_{\kappa\lambda\mu\nu}l^\kappa m^\lambda l^\mu m^\nu,
\end{equation}
\begin{equation}
\Psi_1=C_{\kappa\lambda\mu\nu}l^\kappa k^\lambda l^\mu m^\nu,
\end{equation}
\begin{equation}
\Psi_2=C_{\kappa\lambda\mu\nu}l^\kappa m^\lambda m^{*\mu} k^\nu,
\end{equation}
\begin{equation}
\Psi_3=C_{\kappa\lambda\mu\nu}k^\kappa l^\lambda k^\mu m^{*\nu},
\end{equation}
\begin{equation}
\Psi_4=C_{\kappa\lambda\mu\nu}k^\kappa m^{*\lambda} k^\mu m^{*\nu}.
\end{equation}
\end{subequations}
Therefore, the algebraically complete set of second-order invariants is $\{R, I, I_6, K\}$ \cite{Zakhary:1997acs,Torres:2016pgk,Torres:2022twv}, which means that Eq.~(\ref{eq:rmetric-ss}) is regular if the set of invariants does not diverge anywhere. 
The definitions of $R$ and $K$ have been given in Sec.~\ref{sec:curvatures}, and the definitions of $I$ and $I_6$ are
\begin{equation}
I\equiv\frac{1}{24}C^*_{\alpha\beta\gamma\delta}C^{*\alpha\beta\gamma\delta},
\end{equation}
\begin{equation}
I_6\equiv\frac{1}{12}{\mathcal{S}_\alpha}^\beta{\mathcal{S}_\beta}^\alpha,
\end{equation}
which also belong to the seventeen ZM invairants.
According to the set of invariants, one can deduce the necessary and sufficient condition for the regularity of Eq.~\eqref{eq:rmetric-ss}: If $\sigma(r)$ is a $C^3$ function, then it demands
\begin{equation}
\sigma(0)=0, \qquad \sigma'(0)=0, \qquad \sigma''(0)=0.
\end{equation}

For the seed metric with two shape functions, a general analytical method is still lacking for the regularity conditions of rotating RBHs, and what one can do is to verify the regularity only by calculating $R$ and $K$ in most cases~\cite{Franzin:2021vnj,Mazza:2021rgq,Kumar:2022vfg}.

\section{Interpretation of regular black holes}
\label{sec:interpretation}
A complete RBH theory refers to physical interpretations from either the quantum theory of gravity, such as loop quantum gravity and asymptotic safety method, or the construction of gravitational sources in the context of classical field theory. 
In this section, we explain RBHs from the perspective of coordinate transformations at first, and then we summarize the techniques for constructing gravitational sources for both non-rotating and rotating RBHs. At the end of this section, we give a short discussion on the scalar hair of RBHs because it relates to classical field interpretations.

\subsection{How to understand regular black holes correctly?}
\label{sec:understanding}

\begin{figure}[!ht]
     \centering
     \begin{subfigure}[b]{0.445\textwidth}
         \centering
         \includegraphics[width=\textwidth]{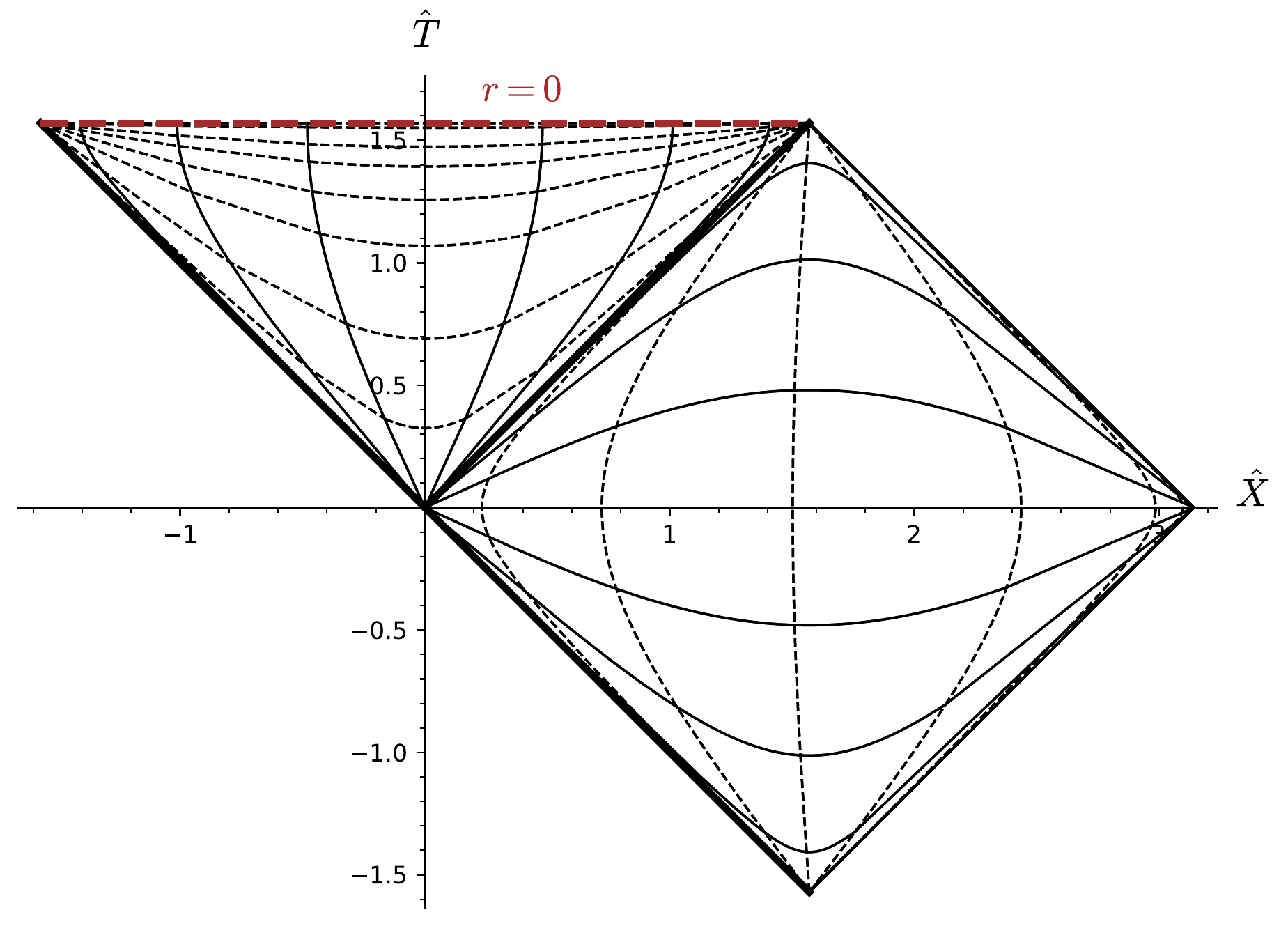}
         \caption{$r\in[0,\infty)$.}
         \label{fig:penrose-original}
     \end{subfigure}
     \begin{subfigure}[b]{0.455\textwidth}
         \centering
         \includegraphics[width=\textwidth]{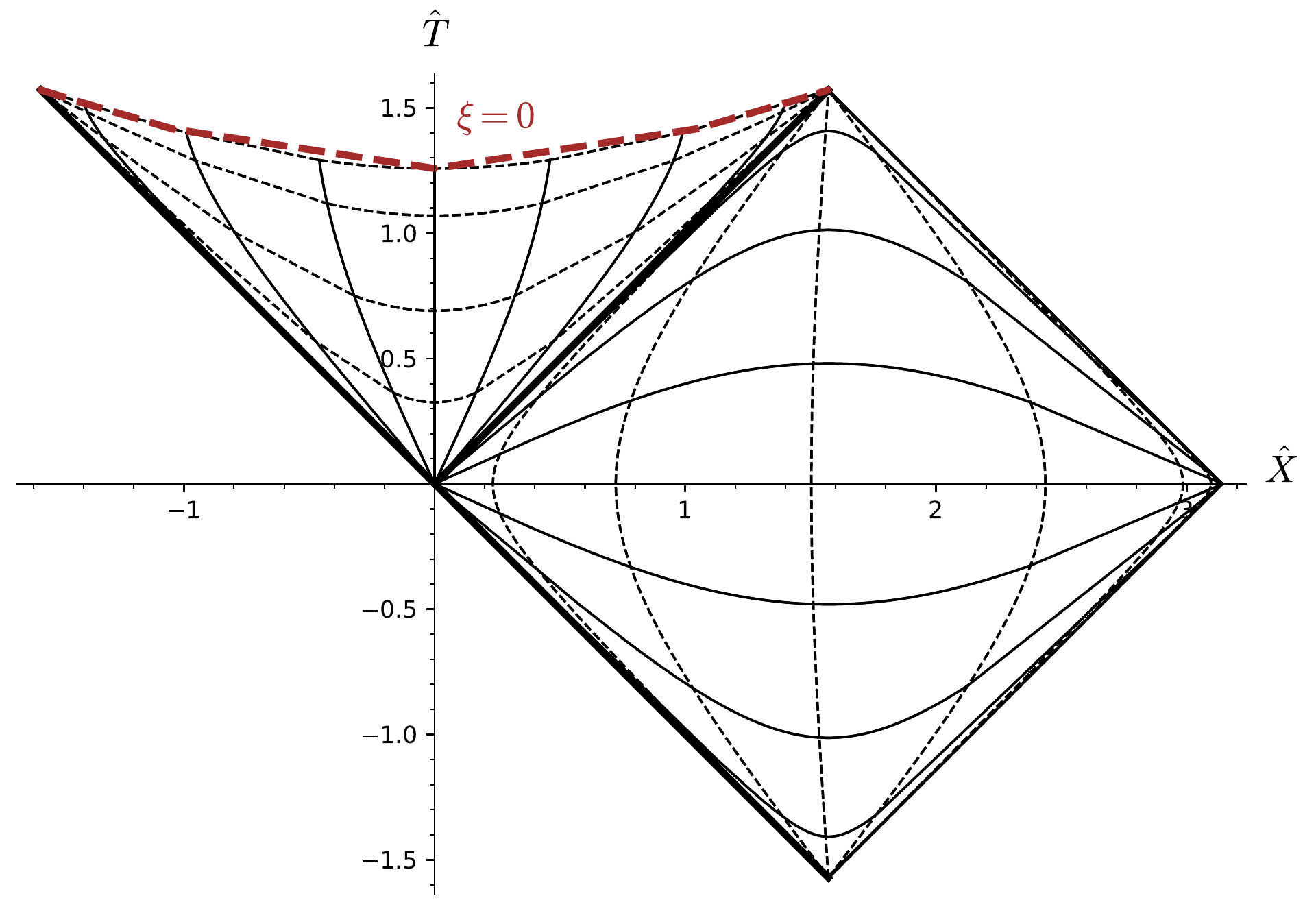}
         \caption{$\xi\in[0,\infty)$}
         \label{fig:penrose-cut}
     \end{subfigure}
      \captionsetup{width=.9\textwidth}
       \caption{Penrose diagrams of a Schwarzschild BH in two different radial coordinates.}
        \label{fig:penrose-schwarzshild}
\end{figure}

It is a confusing issue whether RBHs exist in nature or they are just mathematical tricks, which was emphasized in 
Refs.~\cite{Simpson:2018tsi,Zhou:2022yio}, where a RBH is constructed by a seeming ``coordinate transformation''.
To make the following discussions clear, we ask the question in another way:
\begin{quote}
    \em Is a RBH a full spacetime, or is it just a ``good'' coordinate system that does not cover the entire spacetime in the radial direction?
\end{quote}
Let us examine the Schwarzschild BH as an example to see the essence of the above question, where the Penrose diagram is shown in Fig.\ \ref{fig:penrose-original}.

In order to construct a RBH, we drag the coordinate system downward by a transformation $r\to r(\xi)$, in such a way that the coordinate system cannot cover the singularity after the operation, see Fig.\ \ref{fig:penrose-cut}, where $r\to \sqrt{\xi^2+l^2}$ with $l>0$. In other words, we put the singularity $r=0$ down on the ``non-physical'' domain in the new coordinate system if the new radial coordinate $\xi$ is defined in $\xi\in[0,\infty)$,
i.e., the singularity is dragged to the imaginary axis after analytically continuing  $\xi$ into a complex plane.
This operation is equivalent to restricting the old radial coordinate in $r\in[l,\infty)$ by hands. The singularity is subtracted from the ``old'' spacetime, such that the ``new'' spacetime is regular. In other words, 
the ``new'' coordinates cover a smaller portion of the manifold than the old coordinates in the Schwarzschild spacetime, but the topology of the manifold never changes.
We call the Schwarzschild BH in $\xi$ as a {\em fake} RBH.

Now let us see a
 {\em real} RBH which does not redisplay the singularity by a transformation.
Taking the Bardeen BH as an example \cite{Bardeen:1968nsg}, 
\begin{equation}
\label{eq:bardeen}
    \dif s^2 =-f \dif t^2+ f^{-1} \dif r^2
    +r^2 \dif \Omega^2,\qquad
    f=1-\frac{2M r^2}{(r^2+g^2)^{3/2}},
\end{equation}
where $M$ is mass and $g$ magnetic charge of monopoles, we know that the Kretschmann scalar is regular in $r\in[0,\infty)$.
After a replacement, $r^2\to \xi^2-g^2$, the metric Eq.\ \eqref{eq:bardeen} becomes
\begin{equation}
\label{eq:sing-bardeen}
    \dif s^2 =-f \dif t^2+ \frac{f^{-1} \xi ^2}{\xi ^2-g^2}\dif \xi^2
    +(\xi^2-g^2) \dif \Omega^2,\qquad
    f=1-\frac{2M (\xi^2-g^2)}{\xi^{3}}.
\end{equation}
If we take Eq.\ \eqref{eq:sing-bardeen} as an independent metric describing a ``new'' spacetime, 
the corresponding Kretschmann scalar diverges around $\xi=0$,
\begin{equation}
    K\sim \frac{900 g^8 M^2}{\xi ^{14}}+O\left(\frac{1}{\xi ^{13}}\right).
\end{equation}
It seems that the Bardeen BH redisplays the singularity at $\xi=0$, and
even that almost all the RBHs regain singularities by a replacement of $r\to r(\xi)$.
However, this is not the case because $\xi$ could never be smaller than $g$, otherwise the signature in Eq.\ \eqref{eq:sing-bardeen}  changes in such a way that the line-element represents a manifold with two time dimensions  (breaking of the causality) and two space dimensions and the integral measure $\sqrt{-g}$ becomes complex.

The above two examples reflect the fact that the singularity cannot be resolved by a coordinate transformation, which is consistent with the essence of singularities in BHs.

Further, if one analytically continues the radial coordinate to a complex plane, the complex singularities may emerge again. 
Considering the Kretschmann scalar of Bardeen BHs, 
\begin{equation}
    K=\frac{12 M^2 \left(-4 g^6 r^2+47 g^4 r^4-12 g^2 r^6+8 g^8+4 r^8\right)}{\left(g^2+r^2\right)^7},
\end{equation}
we observe that the singularities are moved to the non-physical domain,  e.g., $r=\pm \mi g\in \mathbb{C}$. 
Generally, one can classify RBHs into three types by the characteristics of singularities \cite{Lan:2022qbb}:
\begin{itemize}
    \item The first type corresponds to those RBHs whose geodesics are complete in the domain of $r\in [0,\infty)$ although their curvature invariants have essential singularities at $r=0$ from the perspective of complex analysis, e.g., $\sigma(r)=\me^{-1/r}$, see Ref. \cite{Balart:2014cga};
    \item The second type corresponds to those RBHs whose singularities of curvature invariants are moved to the non-physical domain, e.g., the Bardeen and Hayward BHs \cite{Hayward:1994bu};
    \item The third type corresponds to those RBHs whose curvature invariants have no singularities on the entire complex plane, e.g., the noncommutative geometry inspired BH \cite{Nicolini:2005vd}.
\end{itemize}
This classification directly affects the calculation of the asymptotic frequencies of quasinormal modes (QNMs) by the monodromy method \cite{Lan:2022qbb}.

\subsection{How to find the sources of non-rotating regular black holes?}
\label{sec:sources}

Gliner discussed~\cite{Gliner:1966} an algebraic property of a four-dimensional energy-momentum tensor (EMT) denoted by $[(1111)]$,
where symbol $1$ corresponds to one diagonal component of the EMT and the parentheses imply equal components,\footnote{As in Ref.\ \cite{Gliner:1966}, we do not distinguish between time and space components by a comma.} see Refs.~\cite{Stephani:2003tm,Petrov:1961pej} for the Segr\'e notations.
The matter with the algebraic property $[(1111)]$, called a $\mu$-vacuum, has a de Sitter-like metric and thus avoids singularities. Later, Gliner's work was extended \cite{Dymnikova:1992ux,Elizalde:2002yz}, where
there are four types of algebraic properties in general for spherically symmetric BHs,
\begin{equation}
    [(1111)],\qquad
    [(11)(11)],\qquad
    [11(11)],\qquad
    [(111)1].
\end{equation}
The matter with these algebraic properties can generate RBHs. 

For instance, one RBH given in Ref.\ \cite{Dymnikova:1992ux} has the property $[(11)(11)]$. Generally, all RBHs with metric Eq.\ \eqref{eq:sphe-met-1}
can have this algebraic property because the Einstein tensor is of the following form,
\begin{subequations}
\begin{equation}
    G^0_{\; 0}=G^1_{\; 1} =\frac{f'(r)}{r}+\frac{f(r)}{r^2}-\frac{1}{r^2},
\end{equation}
\begin{equation}
    G^2_{\; 2}=G^3_{\; 3} =\frac{f'(r)}{r}+\frac{f''(r)}{2},
\end{equation}
\end{subequations}
where we have supposed that Einstein's theory of gravity still holds, $G^\mu_{\;\; \nu}=8\pp T^\mu_{\;\; \nu}$, thus we can discuss the algebraic properties of EMTs in terms of Einstein's tensor.

One example given in Ref.\ \cite{Simpson:2018tsi}
has the property $[11(11)]$.
For the metric with the form
of Eq.\ \eqref{eq:sphe-met-2},
one has
the components of the Einstein tensor,
\begin{subequations}
\label{eq:alg-3}
\begin{equation}
    G^0_{\;\; 0} = \frac{f' \rho '}{\rho }+\frac{f \rho '^2}{\rho ^2}+\frac{2 f \rho ''}{\rho }-\frac{1}{\rho ^2},
\end{equation}
\begin{equation}
     G^1_{\;\; 1} =\frac{f' \rho '}{\rho }+\frac{f \rho '^2}{\rho ^2}-\frac{1}{\rho^2},
\end{equation}
\begin{equation}
G^2_{\;\; 2} =
G^3_{\;\; 3} =
    \frac{f' \rho '}{\rho }+\frac{f''}{2}+\frac{f \rho ''}{\rho },
\end{equation}
\end{subequations}
where $G^0_{\;\; 0}$ does not equal $G^1_{\;\; 1}$ generally.
However, 
when $\rho\propto r$, $G^0_{\;\; 0}=G^1_{\;\; 1}$ appears, and then  $[11(11)]$ reduces to $[(11)(11)]$. 

For the EMT with the algebraic property $[(111)1]$, there is an example in Refs.\ \cite{Bronnikov:2005gm,Bronnikov:2006fu}. Here $[(111)1]$ implies $G^0_{\;\; 0}=G^2_{\;\; 2}=G^3_{\;\; 3}$.

It seems that the algebraic properties do not give us any aid in the construction of RBHs. 
However, these properties are extremely important.
The main reason is that they are associated closely with the construction of RBHs which is different from the construction of SBHs. Let us see the details as follows.

Generally, a complete theory of RBHs can be established with one of two distinct logics. 
The first is the so-called {\em bottom-up} approach, in which the metric with finite curvature invariants is derived based on the First Principle, such as the loop quantum gravity or the theory of asymptotic safety, etc. 
The second logic is the so-called {\em top-down} approach, in which the metric with finite curvature invariants or complete geodesics is postulated at first, and then the classical field that yields such a metric is found.
Therefore, it is necessary to clarify the algebraic properties of the gravitational field for searching matter sources in the second approach.

For instance, the RBH with metric Eq.\ \eqref{eq:sphe-met-1} cannot be interpreted by a scalar phantom field depending only on the radial coordinate, but the RBH with metric Eq.\ \eqref{eq:sphe-met-2} can. 
The reason is that the algebraic property of a scalar phantom field EMT is consistent with the Einstein tensor based on Eq.\ \eqref{eq:sphe-met-2}, i.e., the components of Einstein's tensor match the algebra, $[(111)1]$.

Furthermore, the algebraic properties depend on specific gravitational theories. For the metric Eq.\ \eqref{eq:sphe-met-1}, the algebra is $[(11)(11)]$ in the Einstein gravity, but it changes in the $F(R)$ theory. For instance, if we choose a special case of Starobinsky’s action \cite{Starobinsky:1980te,Vilenkin:1985md,Olmo:2015axa}, 
\begin{equation}
    F(R)=R+\alpha R^2,
\end{equation}
the gravitational equations read
\begin{equation}
F^\mu_{\;\;\nu}\equiv   F'(R) R^\mu_{\;\;\nu}-
    \frac{1}{2}F(R)g^\mu_{\;\;\nu} -[\nabla^\mu\nabla_\nu-g^\mu_{\;\;\nu}\Box]F'(R)=8\pp T^\mu_{\;\;\nu}.
\end{equation}
For the metric Eq.\ \eqref{eq:sphe-met-1}, the components of tensor $F^\mu_{\;\;\nu}$ take the forms,
\begin{subequations}
\begin{equation}
\begin{split}
   2 r^4 F^0_{\;\;0}  =
   & -4 \alpha -2 r^3 f' \left(2 \alpha  f''+\alpha  r f^{(3)}-1\right)\\
   &+2 f \left[12 \alpha -2 \alpha  r^2 \left(r^2 f^{(4)}+2 f''+6 r f^{(3)}\right)
   +8 \alpha  r f'+r^2\right]\\
   &+4 \alpha  r^2 f'^2+\alpha  r^4 f''^2-20 \alpha  f^2-2 r^2,
\end{split}
\end{equation}
\begin{equation}
\begin{split}
    2 r^4 F^1_{\;\;1} = &
    -4 \alpha -2 r^3 f' \left(2 \alpha  f''+\alpha  r f^{(3)}-1\right)\\
    &+2 f \left[-12 \alpha -4 \alpha  r^2 \left(4 f''+r f^{(3)}\right)+8 \alpha  r f'+r^2\right]\\
    &+4 \alpha  r^2 f'^2+\alpha  r^4 f''^2+28 \alpha  f^2-2 r^2,
\end{split}
\end{equation}
\begin{equation}
\begin{split}
    2 r^4 F^2_{\;\;2}= 
    2 r^4 F^3_{\;\;3}=
    &4 \alpha +2 r f' \left[-12 \alpha 
    -2 \alpha  r^2 \left(5 f''+r f^{(3)}\right)+r^2\right]\\
    &+4 \alpha  f \left[2 r^2 f''-r^3\left(5 f^{(3)}+r f^{(4)}\right)+8 r f'+6\right]\\
    &+8 \alpha  r^2 f'^2+r^4 f'' \left(1-\alpha  f''\right)-28 \alpha  f^2.
\end{split}
\end{equation}
\end{subequations}
Generally, we have the algebraic structure, $[11(11)]$. In other words, the RBH with the metric Eq.\ \eqref{eq:sphe-met-1} can be generated by the matter with the algebra, $[11(11)]$. 

Similarly, the change of algebraic structures appears in the conformal gravity with the following action \cite{Bambi:2016wdn},
\begin{equation}
\label{eq:conformal}
   S=\int\dif^4 x\sqrt{-g}\; W,
\end{equation}
where $W$ is the Weyl scalar defined by contracting two Weyl tensors.
The variation to the action gives the gravitation-dependent field, $B^\mu_{\;\;\nu}$, which is called the Bach tensor. 
For the metric Eq.\ \eqref{eq:sphe-met-1}, we obtain
\begin{subequations}
\begin{equation}
\begin{split}
    24 r^4 B^0_{\;\;0}=&
    -4 r f \left[r \left(r^2 f^{(4)}-f''+3 r f^{(3)}\right)+2 f'\right]\\
    &+r^2 \left(r f''-2 f'\right)^2-2 r^4 f^{(3)} f'+4 f^2-4,
\end{split}
\end{equation}
\begin{equation}
    24 r^4 B^1_{\;\;1}=-2 r^3 f^{(3)} \left(r f'-2 f\right)+\left[r \left(r f''-2 f'\right)+2 f\right]^2-4,
\end{equation}
\begin{equation}
\begin{split}
24 r^4 B^2_{\;\;2}=
24 r^4 B^3_{\;\;3}=
    &-r^2 \left(r f''-2 f'\right)^2+2 r^4 f^{(3)} f'-4 f^2+4\\
    &+2 r f \left[r \left(r^2 f^{(4)}-2 f''+2 r f^{(3)}\right)+4 f'\right],
\end{split}
\end{equation}
\end{subequations}
whose algebraic structure is also $[11(11)]$.

\subsection{What are the difficulties for us to find the sources of rotating regular black holes?}

The physical interpretation of a rotating RBH should coincide with that of its non-rotating counterpart called a seed metric.
For the seed metric with one shape function, the physical interpretation contains two aspects: An imperfect fluid and a gravitational field coupled to nonlinear electrodynamics.
Although the interpretation from an imperfect fluid is trivial,
 it is often adopted for spacetime without electromagnetic fields.
Meanwhile, the resulting rotating RBH matches~\cite{Torres:2016pgk,Torres:2022twv,Beltracchi:2021ris} the Segr\'e type, $[(11)(1 1)]$.
So the interpretation works well in the aspect of imperfect fluids for the models in Refs.~\cite{Smailagic:2010nv,Azreg-Ainou:2014aqa}.

For the spacetime with electromagnetic fields, it is widely used~\cite{Ayon-Beato:1998hmi,Ayon-Beato:2000mjt,Balart:2014cga} for the physical interpretation that a gravitational field is coupled to nonlinear electrodynamics.
However, it is difficult to extend this interpretation to a rotating spacetime.
The main reason is that the number of non-zero components of $F_{\mu\nu}$ is changed from one to four by the introduction of rotation, that is, $F_{01}, F_{02}, F_{13}$ and $F_{23}$ are non-trivial, where the field strength is defined by  $F_{\mu\nu}\equiv \partial_\mu A_\nu-\partial_\nu A_\mu$.
In the metric Eq.~\eqref{eq:rmetric-ss}, these four components satisfy~\cite{Dymnikova:2017wlx,Dymnikova:2015hka} the relations,
\begin{equation}
F_{31}=a\sin^2\theta F_{10},\qquad aF_{23}=(r^2+a^2)F_{02}.
\end{equation}
The gravitational field coupled to nonlinear electrodynamics is described by the action,
\begin{equation}
S=\frac{1}{16\pp}\int{\dif^4x\sqrt{-g}[R-\mathscr{L}(F)]},
\end{equation}
\begin{equation}
F=F_{\mu\nu}F^{\mu\nu}.
\end{equation}
Using the Einstein field equations,
\begin{equation}\label{eq:NED-Einstein}
G_{\mu\nu}=2\mathscr{L}_FF_{\mu\alpha}{F_{\nu}}^{\alpha}-\frac{1}{2}\delta_{\mu\nu}\mathscr{L},
\end{equation}
one can determine $\mathscr{L}$ and $\mathscr{L}_F$, where $\mathscr{L}_F\equiv \dif\mathscr{L}/\dif F$.
To further determine $F_{\mu\nu}$ one needs to utilize the dynamic equations.
The variation of the action with respect to $A^\mu$ yields the dynamic equations,
\begin{equation}
\nabla_\mu\left(\mathscr{L}_FF^{\mu\nu}\right)=0,\qquad {\nabla_\mu} ^*F^{\mu\nu}=0,
\end{equation}
where $^*F^{\mu\nu}\equiv \frac{1}{2}\eta^{\mu\nu\alpha\beta}F_{\alpha\beta}$, and  $\eta^{0123}=-1/\sqrt{-g}$.
Then one obtains that the  non-zero components of $F_{\mu\nu}$ satisfy the following equations,
\begin{subequations}
\begin{equation}
\frac{\partial}{\partial r}\left[(r^2+a^2)\sin\theta\mathscr{L}_FF_{10}\right]+\frac{\partial}{\partial\theta}\left[\sin\theta\mathscr{L}_FF_{20}\right]=0,
\end{equation}
\begin{equation}
\frac{\partial}{\partial r}\left[a\sin\theta\mathscr{L}_FF_{10}\right]+\frac{\partial}{\partial\theta}\left[\frac{1}{a\sin\theta}\mathscr{L}_FF_{20}\right]=0,
\end{equation}
\begin{equation}
\frac{\partial}{\partial r}F_{20}-\frac{\partial}{\partial\theta}F_{10}=0,
\end{equation}
\begin{equation}
\frac{\partial}{\partial \theta}\left[a^2\sin^2\theta F_{10}\right]-\frac{\partial}{\partial r}\left[(r^2+a^2)F_{20}\right]=0.
\end{equation}
\end{subequations}
Because $\mathscr{L}_F$ is quite complicated and these equations are highly nonlinear, it is almost impossible to solve these equations directly.
Instead of solving the above equations, one then turns to the nonlinear electromagnetic field by studying~\cite{Benavides-Gallego:2018odl,Breton:2019arv,Toshmatov:2017zpr} the change of gauge fields under the NJA.
When the RN metric is transformed into the KN metric, the gauge potential $A_\mu$ changes~\cite{Erbin:2014aya} in the following way.

In the RN  metric, $A_\mu$ can be written as
\begin{equation}
A_\mu=\frac{q}{r}\delta^u_\mu,
\end{equation}
and its contravariant counterpart takes the form,
\begin{equation}
A^\mu=-\frac{q}{r}\delta^\mu_r=-\frac{q}{r}l^\mu,
\end{equation}
where $l^\mu$ is the tetrad in Eq.~\eqref{eq:tetrad}.
Under the transformation governed by Eqs.~\eqref{eq:conversion rule} and \eqref{eq:transform-delta}, the gauge potential becomes
\begin{equation}
\tilde{A}^\mu=-\frac{qr}{\rho^2}\delta^\mu_r,
\end{equation}
and its 1-form reads
\begin{equation}
\tilde{A}=\frac{qr}{\rho^2}(\dif u-a\sin^2\theta\dif\theta),
\end{equation}
which can be written as 
\begin{equation}
\tilde{A}=\frac{qr}{\rho^2}\left(\dif t-\frac{\rho^2}{\Delta}\dif r-a\sin^2\theta\dif\theta\right),
\end{equation}
due to $\dif u=\dif t-\frac{\rho^2}{\Delta}\dif r$.
Because the factor $\frac{qr}{\Delta}$ depends only on $r$, the term of $\dif r$ can be removed by a gauge transformation and the final formulation  
of the gauge potential can be simplified to be 
\begin{equation}
\tilde{A}=\frac{qr}{\rho^2}\left(\dif t-a\sin^2\theta\dif \phi\right),
\end{equation}
which is just the gauge potential of the KN metric.

However, this method encounters a problem in the NJA, that is, the conversion rule Eq.~\eqref{eq:conversion rule} may not be applicable to the gauge potentials in RBHs.
For example, the gauge field of spherically symmetric RBHs with the magnetic charge $Q_m$ is $A_\mu=Q_m\cos\theta\delta^\phi_\mu$, then the  gauge field of rotating RBHs will become~\cite{Toshmatov:2017zpr}
\begin{equation}
A_\mu=-\frac{Q_m a\cos\theta}{\rho^2}\delta^t_\mu+\frac{Q_m(r^2+a^2)\cos\theta}{\rho^2}\delta^\phi_\mu,\label{amurbhmc}
\end{equation} 
if one uses the above method.
But $\mathscr{L}_F$ calculated by Eq.~(\ref{amurbhmc}) is different~\cite{Rodrigues:2017tfm} from that by Eq.~\eqref{eq:NED-Einstein}, which means that this method should be modified in the case of RBHs.

For the seed metric with two shape functions, there are mainly two types of RBHs: One is associated with the loop quantum gravity and the other type is associated with the black-bounce spacetimes. 
For the first type, the physical interpretation of rotating metrics is just from the loop quantum gravity~\cite{Brahma:2020eos,Kumar:2022vfg}.
And for the second type, the physical interpretation involves two aspects, where one is the stress-energy tensor of a scalar field with nonzero self-interaction potentials and the other is a magnetic field in the framework of nonlinear electrodynamics~\cite{Bronnikov:2022bud}.
In addition, there are two different physical interpretations for rotating black-bounce spacetimes, where one is the gravitational field coupled with a nonlinear electrodynamic field together with a contribution of charged dusts and the other is an anisotropic fluid~\cite{Franzin:2021vnj}. However, it needs further studies to judge
which interpretation is more reasonable.

\subsection{Can regular black holes have scalar hairs?}
\label{sec:scalar}

SBHs are governed by the non-scalar-hair theorem \cite{Herdeiro:2015waa,Khlopov:1985jw}.
In RBHs, the situation is improved \cite{Bronnikov:2005gm,Bambi:2016wdn,Kamenshchik:2023bzc,Karakasis:2023hni}. For instance, the metric of conformal RBHs in Ref.
\cite{Bambi:2016wdn} reads
\begin{equation}
    \dif s^2 = \left(1+\frac{L^2}{r^2}\right)^{2n}
    \left(
    -f\dif t^2
    +f^{-1} \dif r^2
    +r^2 \dif \Omega^2
    \right),
\end{equation}
where $L$ is the regularization parameter with the length dimension and $f=1-2M/r$. This RBH model can be produced by the following action,
\begin{equation}
\label{eq:action-conformal}
I_{\rm conf}=-\frac{1}{2}\int \dif^4 x \sqrt{-g} \phi \left(
\frac{1}{6}R \phi -\Box \phi
\right),
\end{equation}
where $\phi$ is the scalar field.
The equation of motion for the scalar field $\phi$ is
\begin{equation}
    \Box \phi = \frac{1}{6} R \phi,
\end{equation}
and its solution takes \cite{Lan:2021klp} the form,
\begin{equation}
\label{eq:sol-scalar}
\phi =\left(1+\frac{L^2}{r^2}\right)^{-n}\left[\frac{c_1 }{2 M}\ln \left(1-\frac{2 M}{r}\right)+c_2\right],
\end{equation}
where $c_1$ and $c_2$ are integration constants.
This solution is divergent at the horizon, $r_{\rm H} =2M$, which implies $c_1=0$, i.e., the solution can be simplified to be
\begin{equation}
    \phi =c_2\left(1+\frac{L^2}{r^2}\right)^{-n}.
\end{equation}
Since $n\ge 1$, $\phi$ is bounded by $0\le\phi\le c_2$. In other words, Eq.\ \eqref{eq:action-conformal} has a non-trivial scalar hair because of a non-minimal coupling \cite{Herdeiro:2015waa}. 

Another example is shown in Ref.\ \cite{Bronnikov:2005gm,Bronnikov:2006fu}, where the action is cast in the Einstein gravity with a minimally coupled (phantom) scalar field $\phi$,
\begin{equation}
    I_{\rm phatom}=\int\dif^4 x\sqrt{-g} 
    \left[
    R-\partial_\mu\phi \partial^\mu\phi-2 V(\phi)
    \right],
\end{equation}
where $V(\phi)$ is potential, see Ref.\ \cite{Bronnikov:2005gm,Bronnikov:2006fu} for its formulation. This model gives a RBH solution that has the form of Eq.\ \eqref{eq:sphe-met-2} with two shape functions, where one of the functions reads
\begin{equation}
    f(\rho)=1-\frac{\rho _0 \left(\pp  b^2-2 b \rho +\pp  \rho ^2\right)}{2 b^3}+\frac{\rho _0 \left(b^2+\rho ^2\right) }{b^3}\tan^{-1}\left(\frac{\rho }{b}\right),
\end{equation}
where we have used the condition $2 b c=-\pp \rho_0$ with $\rho_0>0$ and $b>0$ to replace $c$ in the original formula in Ref.\ \cite{Bronnikov:2005gm,Bronnikov:2006fu}. 
In addition,
the potential takes the form,
\begin{equation}
    V(\phi)=-\frac{\rho _0}{2 b^3} \left[2 \sqrt{2} \phi +3 \sin (\sqrt{2} \phi)+(\pp -\sqrt{2} \phi) \cos (\sqrt{2} \phi)-2 \pp \right],
\end{equation}
thus we obtain
\begin{equation}
    \phi V'(\phi)=\frac{\rho _0 \phi  }{2 b^3}
    \left[(\sqrt{2} \pp -2 \phi) \sin (\sqrt{2} \phi)-2 \sqrt{2}\cos (1+\sqrt{2} \phi)\right].
\end{equation}
Because $\phi V'(\phi)$
is not always positive, it is not restricted \cite{Herdeiro:2015waa} by the no-hair theorem. Thus the non-trivial scalar-hair solution exists,
\begin{equation}
    \phi=\pm \sqrt{2}\tan^{-1}\left(\frac{\rho}{b}\right)
+\phi_0,
\end{equation}
where $\phi_0$ is integration constant, and it is bounded by $\phi_0-\pp/\sqrt{2}<\phi<\phi_0+\pp/\sqrt{2}$.

\section{Energy conditions of regular black holes}
\label{sec:energyconditions}

The energy conditions are important to the study of RBHs. On one hand, they are related to the formation of RBHs, and on the other hand, they are regarded as criteria if a RBH is realistic or not. In this section, we explain these two aspects.

\subsection{Is the strong energy condition a key to lead to a regular black hole?}
\label{sec:strong}

It was originally thought \cite{Dymnikova:1992ux,Dymnikova:2015yma} that RBHs can be constructed when the singularity at their centers is replaced by a de Sitter core, which implies the violation of the strong energy condition (SEC). 

Because of this, RBHs are not governed by the Penrose singularity theorem, which 
can be understood from the Raychaudhuri equation \cite{Carroll:2004st,Poisson:2009pwt},
\begin{equation}
\label{eq:raych}
\frac{\dif\Theta}{\dif \tau}=
-R_{\mu\nu} u^\mu u^\nu,
\end{equation}
where $\tau$ is proper time, $u^\mu$ is four-velocity and $\Theta$ depicts the expansion of geodesic congruence. For simplicity, we have already ignored higher order terms associated with expansion, rotation, and shear in the right hand of Eq.\ \eqref{eq:raych}. 
Then, choosing $u^\mu=(1,0,0,0)$,
we arrive at 
\begin{equation}
    \frac{\dif\Theta}{\dif \tau}=-R_{00}=-4\pp G \left(\rho+\sum_{i=1}^3 p_i\right),
\end{equation}
where $\rho$ is energy density and $p_i$ are three components of pressure. 
The violation of SEC, $\rho+\sum_{i=1}^3 p_i<0$, implies an increasing $\Theta$ along the proper time, i.e., the interaction is repulsive.

Nevertheless, it was also discovered \cite{Balakin:2015gpq,Balart:2014cga,Simpson:2019mud,Li:2021bmj} that RBHs can have an anti-de Sitter or a flat core. For instance, the RBH constructed in Ref.\ \cite{Balakin:2015gpq} is described by
\begin{equation}
\label{eq:balakin}
    \dif s^2 =-f \dif t^2+ f^{-1} \dif r^2
    +r^2 \dif \Omega^2,\qquad
    f=1+\frac{r^4 }{r^4+2 q Q_m^2}\left(-\frac{2 M}{r}+\frac{Q_m^2}{r^2}\right),
\end{equation}
where $q$ is a {\em positive} parameter describing the non-minimal coupling of Yang-Mills fields, $Q_m$ is a magnetic charge, and the cosmological constant is set to be zero in order to highlight the essence. 
The anti-de Sitter core can be seen clearly from the following asymptotic relations,
\begin{equation}
    f\sim 1+\frac{r^2}{2q}+O(r^3),\qquad
    R\sim -\frac{6}{q}+O\left(r\right)
\end{equation}
as $r$ approaches zero. Moreover, a spherically symmetric RBH model with a flat core is highlighted in Ref.\ \cite{Simpson:2019mud,Ling:2021olm},
where the shape function reads \cite{Culetu:2014lca} 
\begin{equation}
\label{eq:balart}
    f=1-\frac{2M }{r}\me^{-a/r}.
\end{equation}
It is obvious that $f\to 1$ and $R\to 0$ as $r$ approaches zero because the parameter $a$ is positive.
These two examples meet the SEC in the cores because the AdS and Minkowski spacetimes satisfy the SEC.

The problem immediately arises, if the SEC does not break, i.e., the gravity is attractive in the core, how the collapse can be avoided. 
One interesting resolution is based \cite{Zaslavskii:2010qz} on the introduction of the Tolman mass which can be regarded as a kind of {\em integral} SEC \cite{Tolman:1987rtc,Abreu:2010sc},
\begin{equation}
    m_{\rm T}=\frac{1}{4\pp}\int\sqrt{-g}R_{00}\dif^3 x=
    \int r^2 R_{00}\dif r.
\end{equation}
The {\em integral} SEC breaks in the core ($r\in[0,r_-]$, $r_-$ is the innermost horizon) if the Tolman mass is negative. 
Due to the negative Tolman mass in these two models described by Eq.\ \eqref{eq:balakin} and Eq.\ \eqref{eq:balart}, we can conclude that the two models violate the integral SEC in their cores. 

In summary,\footnote{According to Ref.\ \cite{Curiel:2014zba},
the energy conditions can be divided into two categories: One restricts average behaviors across regions of spacetime, and the other restricts behaviors at specific points. Here the ``impressionist SEC" means the former, and the ``pointillist SEC" means the latter.} it is the impressionist SEC that plays the key role in constructing RBHs, not the pointillist one \cite{Curiel:2014zba}. Because of this, it is not important whether the cores of RBHs are de Sitter, anti-de Sitter or flat although the violation of SEC is a necessary condition for the formation of RBHs from gravitational collapse  \cite{Zhang:2014bea}.

\subsection{What are the energy conditions of regular black holes?}
\label{sec:engcond}

If the SEC answers how RBHs are formed,
then the other three energy conditions actually focus on whether RBHs are realistic  \cite{Mars:1996khm,
	Borde:1996df,
	Aftergood:2014wla,
	Balart:2014jia,
	Rodrigues:2017yry,Dymnikova:2017wlx} through the characteristics of {\em classical} matters.

Besides the SEC, the other three energy conditions are the  
weak energy condition (WEC), the null energy condition (NEC), and the dominant energy condition (DEC) \cite{Curiel:2014zba}. 
The violation of WEC implies a negative local energy density, and the violation of DEC means the superluminal speed of energy density flow. Moreover, the NEC is a relaxation of WEC, i.e., the energy density can be negative as long as the sum of energy density and pressure is positive \cite{Carroll:2004st}.

For the RBHs with one shape function, see Eq.\  \eqref{eq:shape1}, the other three energy conditions except the SEC can be reduced to the following differential inequalities,
\begin{equation}
\label{eq:ECDs}
    \begin{split}
    \text{WEC}&:\quad  \sigma '\ge 0\;\cup\; r \sigma ''\le 2 \sigma ',  \\
    \text{NEC}&:\quad  r \sigma ''\le 2 \sigma',  \\
     \text{DEC}&:\quad \sigma '\ge 0\;\cup\; -2 \sigma'\le r   \sigma ''\le 2 \sigma',
    \end{split}
\end{equation}
where the prime denotes the derivatives w.r.t.\ $r$. The relation between these three energy conditions can be represented by
\begin{equation}
    \text{NEC} \subset
    \text{WEC} \subset
    \text{DEC}.
\end{equation}
Among these three independent differential inequalities in Eq.\ \eqref{eq:ECDs}, $\sigma'\ge 0$ implies that $\sigma$ is a monotone increasing function
of $r$; while $r \sigma'' \le 2 \sigma'$ provides $\sigma\le \sigma_0 r^3$, where $\sigma_0\equiv \lim_{r\to 0} \sigma/r^3$; the last one $r \sigma''\ge-2 \sigma'$ gives a solution, $r \sigma\ge 0$, under the boundary conditions, $\sigma|_{r=0}=0=\sigma'|_{r=0}$.

For the RBHs with two shape functions, see Eq.\ \eqref{eq:sphe-met-2}, the situation becomes complicated. The differential inequalities now involve an additional unknown function, $r(\xi)$, which causes the inequalities to be unsolvable without extra specific constraints. Therefore, we cannot extract any valuable information from these differential inequalities.

In addition, some RBH models violate \cite{Toshmatov:2017kmw,Maeda:2021jdc,Lan:2022qbb} the three energy conditions. 
For instance, the well-known Bardeen and Hayward BHs break the DEC, and 
the RBH generated by the non-minimal coupled Wu-Yang monopole brakes \cite{Balakin:2006gv,Balakin:2016mnn,Liu:2019pov,Rayimbaev:2021vsq,Jusufi:2020odz} the WEC, etc. 
A remedy is proposed in Ref.\ \cite{Lan:2022bld} by deforming the shape function.
For a generic $\sigma$ function, 
its deformed formulation reads
\begin{equation}
\label{eq:deform-sigma}
\sigma=  \frac{M^{\mu\nu-3} r^3}{\left(r^{\mu }+q^{\mu }\right)^{\nu}},
\end{equation}
where $M$ is mass and $q$ regularization parameter, which 
 contains the Bardeen and Hayward BHs as special cases. 
Meanwhile, it
will meet the three energy conditions
if the parameters $\mu$ and $\nu$ take the values in the following regions,
\begin{subequations}
\label{eq:dec-generic}
\begin{eqnarray}
\frac{2}{\nu }<\mu \leq \frac{1}{2} \sqrt{\frac{49 \nu +96}{\nu }}-\frac{7}{2}
&&\quad \text{when}\quad \frac{2}{5}<\nu \leq 3;
\\
\frac{2}{\nu }<\mu \leq \frac{3}{\nu }
&&\quad \text{when}\quad \nu>3,
\end{eqnarray}
\end{subequations}
where $M^{\mu\nu-3}$ is introduced for balancing the dimension, and this parameterization is not unique.

\section{Thermodynamics of regular black holes}
\label{sec:thermodynamics}

The thermodynamics of RBHs is a rather confusing battleground in the study of RBHs. The confusion comes from the existence of extra terms in the first law of {\em mechanics} such that the correspondence between mechanical and thermodynamic quantities is problematic. In this section, we give our clarifications on these problems in the context of Einstein's gravity coupled with nonlinear electrodynamics.

\subsection{What is the entropy of regular black holes?}
\label{sec:entropy}

It was reported \cite{Myung:2007qt,Myung:2008kp,Spallucci:2008ez,Miao:2015npc,Nam:2018sii,Lan:2020wpv,Kumara:2020ucr} that RBHs have a deviation term in entropy, i.e., the entropy of RBHs breaks the area law, $S\neq A/4$, but the opposite opinion was also reported \cite{Banerjee:2008gc,Fan:2016hvf,Fan:2016rih,Kruglov:2017fck,Nojiri:2017kex,Guo:2021wcf}.
This puzzle gives rise to an influence on the first law of thermodynamics and the interpretation of RBHs,
where ambiguous deviation terms appear for the former, while the verification of Hawking's quantum theory is hard to be executed for the latter.

Taking the Hayward BH \cite{Hayward:2005gi} as an example, whose shape function is
\begin{equation}
    f=1-\frac{2M}{r}
    \frac{r^3}{r^3+2 Ml^2},\label{eq:shafunhaybh}
\end{equation}
where $l$ is the length scale introduced for regularization, we can obtain the entropy from the first law of thermodynamics, $\dif M = T \dif S$, 
\begin{equation}
\label{eq:entropy-therm}
      S=
   \int^{r_{+}}_{r_{-}} \frac{\dif M}{T}
   =S_{\rm BH}+\Delta S,
\end{equation}
where $r_+$ and $r_-$ are outer and inner horizons, respectively, $S_{\rm BH}$ is the Bekenstein-Hawking entropy, 
\begin{subequations}
\begin{equation}
    S_{\rm BH}=\pp  \left(r_+^2-r_-^2\right),
\end{equation} 
and $\Delta S$ is a deviation term,
\begin{equation}
    \Delta S= \frac{\pp  l^4\left(r_+^2-r_-^2\right)}{\left(r_-^2-l^2\right) \left(r_+^2-l^2\right)}+
    2 \pp  l^2 \ln \left(\frac{r_+^2-l^2}{r_-^2-l^2}\right).
\end{equation}
\end{subequations}
We note $\Delta S>0$ because of $r_+\ge r_->l$, otherwise, there will be no horizons.
As a result, if Eq.\ \eqref{eq:entropy-therm} is still used to calculate the entropy, the area law is violated, $S\neq A/4$, even though a pressure term,
\begin{equation}
    \mathcal{P}=-\frac{3}{8\pp l^2},\label{eq:predsst}
\end{equation}
is added  \cite{Sekiwa:2006qj,Wang:2006bn}. 
We note that the pressure in Eq.\ \eqref{eq:predsst} is introduced in a dS spacetime, not in an AdS spacetime due to the minus sign meaning an outward pressure from the center of BHs. For a varying cosmological constant, its corresponding pressure can be introduced whether a spacetime is an AdS or a dS one because the thermodynamic relations of BHs can remain consistent from a mathematical standpoint, see, for instance, Refs.
 \cite{Sekiwa:2006qj,Wang:2006bn,Kastor:2009wy}. 
Here the Hayward BHs have a dS core. On one hand, the dS core is related to a length scale regarded as the regularization parameter in Eq.\ \eqref{eq:shafunhaybh}. On the other hand, it is related to the pressure expressed by the so-called dS radius in Eq.\ \eqref{eq:predsst}. Considering the uniqueness of the cosmological constant, 
we can deduce the equality of the two length scales in Eqs.\ \eqref{eq:shafunhaybh} and \eqref{eq:predsst}.

As a matter of fact, 
if we interpret the metric, Eq.\ \eqref{eq:sphe-met-1}, as a spacetime produced by a field of Dirac magnetic monopoles \cite{Ayon-Beato:1998hmi,Ayon-Beato:2004ywd}, 
the entropy can be derived by Hawking's path-integral method for a given RBH depicted by Eq.\ \eqref{eq:sphe-met-1}.

To start with, we write down 
the full action,
\begin{equation}
\label{eq:action-em}
I=\frac{1}{16\pp} \int \dif^4 x \sqrt{-g}\, \left[R-\ml(F)\right],
\end{equation}
where 
$F$ is the contraction of the electromagnetic tensors, $F\equiv F_{\mu\nu}F^{\mu\nu}=2q/r^4$, and $q$ is the magnetic charge of the monopole.
By solving one of Einstein's equations \cite{Fan:2016hvf},
\begin{equation}
\label{eq:lagrange}
\frac{\mathcal{L}}{2}-\frac{2 M \sigma '(r)}{r^2}=0,
\end{equation}
and replacing $r$ by $r=[F/(2 q)]^{1/4}$, we can determine the Lagrangian $\ml(F)$.

To calculate the entropy, 
we  apply the path-integral method in the zero-loop approximation \cite{Gibbons:1976ue},
\begin{equation}
{Z}=\int {D} g\,{D} A\; \me^{-I}\approx  \me^{-I_{\rm cl}},
\end{equation}
where the full Euclidean action $I_{\rm cl}$ consists of four parts, 
\begin{equation}
\label{eq:class-action}
I_{\rm cl} = I_{\rm EH}+I_{\rm GHY}-I_0+I_M.
\end{equation}
$I_{\rm EH}$ is the Einstein-Hilbert action,
\begin{equation}
\label{eq:action-eh}
I_{\rm EH} =-\frac{1}{16\pp}\int_{\mathcal{M}}\dif^{4}x\sqrt{-g}\,R,
\end{equation}
$I_{\rm GHY}$ is the Gibbons-Hawking-York boundary term, 
\begin{equation}
	I_{\rm GHY} =-\frac{1}{8\pp} \int_{\partial\mathcal{M}}\dif^{3}x\sqrt{-h}\, K,
\end{equation}
$I_0$ is the subtraction term,  
\begin{equation}
	\label{eq:counterterm}
	I_0 =-\frac{1}{8\pp} \int_{\partial\mathcal{M}}\dif^{3}x\sqrt{-h}\, K_0,
\end{equation}
and $I_M$ is the matter action of a nonlinear electrodynamic source,
where $R$ is the Ricci curvature of bulk space, $K$ and $K_0$ are extrinsic curvatures of surface
and background reference, respectively. 

For the metric Eq.\ \eqref{eq:sphe-met-1}, we obtain
\begin{gather} 
I_{\rm EH}= 
\frac{\beta}{2}   \left(r_{\rm H}-M\right)-\pp  r_{\rm H}^2,\label{eqahaction}\\
I_{\rm GHY} =
\frac{M }{2}\beta\left(r \sigma'+3 \sigma\right)
-r \beta\sim
\frac{3M }{2}\beta
-r \beta,\\
I_0 =-r\beta\sqrt{1-\frac{2 M \sigma}{r}}
\sim -r\beta+M\beta+O\left(r^{-1}\right),
\end{gather}
where $\beta$ is the inverse of BH temperature, $1/T$, and the last two asymptotic relations at the boundary $r\to \infty$ are derived based on the assumption
that the RBH is asymptotic to the Schwarzschild BH at infinity \cite{Balart:2014cga}, i.e.,
$\lim_{r\to\infty}\sigma =1$
and
$\lim_{r\to\infty} r\sigma' =0$. 
Meanwhile,
the action of matter can be determined with the help of one component of Einstein's equations Eq.\ \eqref{eq:lagrange},
\begin{equation}
I_M=\beta  M-\frac{\beta  r_{\rm H}}{2}.\label{eqmattaction}
\end{equation}
Substituting Eqs.\ \eqref{eqahaction}-\eqref{eqmattaction} into Eq.\ \eqref{eq:class-action}, we arrive at the total Euclidean action,
\begin{equation}
\label{eq:action-result}
I_{\rm cl} =\beta  M -\pp r_{\rm H}^2.
\end{equation}
On the other hand, 
the thermodynamic law for the canonical ensemble is $\ff=M - T S$, 
where $\ff=T I_{\rm cl}$ is Helmholtz free energy.
Thus, we can read $S=\pp r_{\rm H}^2$ from Eq.\ \eqref{eq:action-result}, which
exhibits the well-known entropy-area law of BHs.
The same result can be obtained by Wald's Noether-charge approach.
Since the Lagrangian of the model, Eq.~\eqref{eq:sphe-met-1} is just $R$, 
the entropy density can be read off directly from Table I in Ref.~\cite{Jacobson:1993vj}.

However, the problem is far more complicated than it appears --- the above calculation relies on the interpretation of the metric. That is, the calculation depends on the nonlinear magnetic interpretation of metrics, see Eq.\ \eqref{eq:action-em}. 
If we reinterpret the source as dyons,
the path-integral method is not applicable because no actions for dyons can be constructed \cite{Fan:2016hvf}.
Furthermore, if we interpret a metric by using an alternative gravity, such as $f(R)$, 
the entropy-area law will be changed.

\subsection{What is the correct first law of thermodynamics for regular black holes?}
\label{sec:themlaw}

One feature of RBHs is that their {\em mechanical} theorems differ from those of SBHs \cite{Zhang:2016ilt}. 
More precisely, unexpected extra terms\footnote{Extra terms indicate additional terms for a parameter, e.g., for mass $M$ in Eq.\ \eqref{eq:first-law}, $K_M \dif M$ is an extra term, while for charge $q$, $K_q\dif q$ is an extra term.} will appear in the first {\em mechanical} laws of RBHs, and the number of extra terms depends on the number of parameters involved in Lagrangian of matters.

For instance, the Lagrangian of Bardeen BHs contains two parameters, mass $M$ and magnetic charge $q$ \cite{Ayon-Beato:2000mjt}, thus the first {\em mechanical} law reads \cite{Zhang:2016ilt}
\begin{equation}
\label{eq:first-law}
\dif M = \frac{\kappa}{8\pp} \dif A +\Psi_H \dif q
+K_M \dif M +K_q \dif q,
\end{equation}
where $\kappa$ is the surface gravity, $\Psi_H$ is the magnetic potential and the last two terms are extra.
As a result, one encounters several problems when attempting to construct the first {\em thermodynamic} law, e.g.,
what is the correspondence between mechanical and thermodynamic variables and what is the dimension of the thermodynamic phase space?

To construct the first {\em thermodynamic} law, Fan and Wang \cite{Fan:2016hvf} introduced a parameter $\alpha$ in the action of nonlinear electrodynamics.
According to their method, 
the first thermodynamic law for the Bardeen BH,  Eq.\ \eqref{eq:bardeen}, can be cast in the following form,
\begin{equation}
\label{eq:fan-wang}
    \dif E = T \dif S +\Psi_{H}\dif Q_m +\Pi \dif \alpha,
\end{equation}
where the three thermodynamic variables,
\begin{equation}
\label{eq:tangle}
    E=M,\qquad Q_m=\sqrt{M q/2}, \qquad
    \alpha=q^3/M=8 Q^6_m/M^4,
\end{equation}
are not independent of each other in the phase space. 
In other words, the dimension of thermodynamic phase space is two because $\alpha$ is a redundant dimension.
The other correspondences between thermodynamic and mechanical variables are
\begin{equation}
    T\longleftrightarrow\frac{\kappa}{2\pp},\qquad
    S\longleftrightarrow\frac{A}{4}.
\end{equation}
Nevertheless, Eq.\ \eqref{eq:fan-wang} is still problematic because
\begin{eqnarray}
\frac{A}{4}=S\neq \int \frac{\dif E}{T}=\int\frac{\dif M}{T},
\end{eqnarray}
where the integral is calculated under the condition, $\dif Q_m=0=\dif \alpha$. 
The reason of $S\neq \int \dif M/T$ or $T\neq (\partial E/\partial S)_{Q,\alpha}$ is that the relations, $\dif Q_m=0=\dif \alpha$, imply that $M$ is also a constant, see Eq.\ \eqref{eq:tangle}, which leads to a trivial integral, $\int\dif M/T\equiv 0$.

If only one parameter is fixed, say $\dif Q_m=0$, we have
\begin{equation}
    \frac{A}{4}=
    \int\frac{\dif M}{T}\left(1+\frac{32 Q_m^6\Pi}{M^5}\right)\neq \int \frac{\dif M}{T},
\end{equation}
namely, the simple thermodynamic relation, $S=\int \dif M/T$, is broken. This result gives rise to a tendency to abandon \cite{Myung:2007qt,Myung:2008kp,Spallucci:2008ez,Miao:2015npc,Nam:2018sii,Lan:2020wpv,Kumara:2020ucr,Tzikas:2018cvs,Singh:2020rnm,Rizwan:2020bhp} the area-entropy law,
\begin{eqnarray}
S=\int \frac{\dif E}{T}\neq \frac{A}{4}.
\end{eqnarray}
In the above treatment, the entropy $S$
cannot be determined if the first law of thermodynamics is not applied. The worst point is that the broken area-entropy relation contradicts the results obtained from either Hawking's path-integral or Wald's entropy formula.

Therefore, the question naturally arises --- what a correct first law of thermodynamics is. We list its most important features below:
\begin{enumerate}
    \item The area law should be maintained, i.e., $S=A/4$, if one explains RBHs in the context of Einstein’s gravity. Generally, the entropy in the first thermodynamic law should be consistent with that calculated from either Hawking's path-integral or Wald's entropy formula.
    \item Every thermodynamic variable should be independent of the first thermodynamic law, i.e., it can be determined without the first thermodynamic law, but the thermodynamic formula, $S=\int \dif E/T$, must hold.
    Here are some counterexamples: Temperature is not independent in Ref.\ \cite{Ma:2014qma}, deviation of internal energy is not independent in Ref.\ \cite{Maluf:2022jjc}, etc.
    \item Every thermodynamic variable should be  independent of each other, e.g., $\dif M = T\dif S + K_1 \dif \alpha + K_2 \dif \beta+\ldots$,
     is ill-defined if $\alpha=M$ and $\beta=T M$ that make
     $\alpha$ and $\beta$ dependent on $M$ in the thermodynamic phase space. 
\end{enumerate}

In order to establish a well-defined first thermodynamic law that meets the above conditions for RBHs with two parameters, such as Bardeen BHs, we have applied \cite{Azreg-Ainou:2014twa,Ma:2014qma,Lan:2021ngq} the following form in terms of the Gliner vacuum,
\begin{equation}
\dif U = T\dif S -P_{+} \dif V,
\end{equation}
where the total internal energy reads
 \begin{equation}
 U=\frac{r_+}{2},
\end{equation}
$P_+$ and $V$ are the thermodynamic pressure and volume, respectively,
\begin{equation}
P_{+}=\left.\frac{G^r_{\; r}}{8\pp}\right|_{r=r_{+}},
\qquad
V=\frac{4}{3}\pp r_+^3.
\end{equation}

\section{Thermodynamic geometry of regular black holes}
\label{sec:bhchemandruppgeo}
Thermodynamic geometry gives a different way to understand BHs and it is also a very powerful tool for exploring the microstructures of RBHs. The study of Ruppeiner geometry and Weinhold geometry can help us to further understand the thermodynamic properties of RBHs.

\subsection{What is thermodynamics of regular black hole?}
\label{sec:bhchem}
\subsubsection{Thermodynamic phase transition and shift of critical points}

In recent years, there have been a lot of studies on
thermodynamic behaviors of  RBHs, and these behaviors are similar to those in thermodynamics of SBHs,
such as gas/liquid phase transitions \cite{Tzikas:2018cvs,Liu:2022spy,NaveenaKumara:2019nnt}, van der Waals-like fluid properties \cite{Tharanath:2014naa,Molina:2021hgx,Rodrigues:2022qdp,Li:2018bny},
triple points \cite{NaveenaKumara:2020jmk,Rodrigue:2018lzp}, and heat engines \cite{Rajani:2019ovp,Guo:2021bju,Guo:2019rdk,Ye:2020aln,Nam:2019zyk}.
For a given equation of state, the critical point ($T_c$, $P_c$, $V_c$) of phase transitions can be determined by the following condition, 
\begin{equation}
	\label{pt}
	\left( \frac{\partial P}{\partial V}\right)_T=0, \qquad \left( \frac{\partial^2 P}{\partial V^2}\right)_T=0.
\end{equation}
The behaviors of critical points allow us to make a deep analogy between BHs and gas/liquid systems.  
In particular, the phase transitions of van der Waals-like fluid have been well tested \cite{Zhang:2021raw,Kumara:2021hlt} in different theories of gravity. 
However, due to the problematic first law of thermodynamics, the Maxwell equal area law is no longer valid for RBHs. The exotic equal area law for Hayward AdS BHs reads~\cite{Fan:2016rih}
\begin{equation}
	\oint VdP=-\oint\Psi dQ_m+\oint \text{extra term},
\end{equation}
which makes Maxwell's equal area law invalid in the $P-V$ plane. Thus, the critical point ($T_c$ , $P_c$) of the first-order small/large BH transition does not coincide with the inflection point of isotherms.

\subsubsection{Regular black hole as a heat engine}
That BHs behave like the van der Waals fluid and have gas/liquid phase transitions also makes it possible to treat BHs as heat engines. The first holographic heat engine was proposed \cite{Johnson:2014yja} by Johnson, where BHs act as a working substance. The heat engine flows
in a cycle as shown in Fig. 1 of Ref.~\cite{Johnson:2014yja}.
Using equations of state of AdS BHs, for instance, we can compute the total mechanical work as follows,
\begin{equation}
	W=Q_H-Q_C,
\end{equation}
where $Q_H$ is the net input heat flow and $Q_C$ is the net output heat flow.
Therefore, the efficiency of heat engines is defined by
\begin{equation}
	\eta\equiv\frac{W}{Q_H}=1-\frac{Q_C}{Q_H}.
\end{equation}

Recently, some progress has been made \cite{Johnson:2015ekr,Rajani:2019ovp,Guo:2019rdk,Sharma:2022fjh} in the studies of heat engines for RBHs, where the relation between efficiency and entropy (pressure) has been obtained \cite{Guo:2019rdk} and the comparisons of efficiency among RBHs have also been made. 
We note that it is subtle to construct heat engines and especially the engine cycles in the $P-V$ plane for RBHs, and that some other progress has provided~\cite{Caceres:2015vsa,Mo:2017nhw,Johnson:2016pfa} a new perspective on thermodynamic properties of RBHs.

\subsection{How to eliminate the singularity of thermodynamic geometry for regular black holes?}
\label{sec:ruppgeo}
\subsubsection{Construction of thermodynamic geometry}
A similar situation has happened in the studies of 
thermodynamic geometry, where such geometry has been shown to be a powerful tool for understanding the thermodynamic properties and microstructures of SBHs.
The Ruppeiner metric is defined~\cite{Ruppeiner:1995zz} by the Hessian matrix of thermodynamic entropy,
\begin{equation}
	g^{\rm R}_{\mu\nu}\equiv -\frac{\partial^2S(X)}{\partial X^\mu \partial X^ \nu},
\end{equation}
and the Weinhold metric is defined~\cite{weinhold1975metric} by the Hessian matrix of BH mass,
\begin{equation}
	g^{\rm W}_{\mu\nu}=\frac{\partial^2M(X)}{\partial X^\mu \partial X^ \nu},
\end{equation}
where $X^\mu$ denote thermodynamic variables.
Further, we can see that there is the following conformal relationship between the two kinds of  thermodynamic geometries if we write the metrics in terms of line elements,
\begin{equation}
	\label{Conformal}
	\dif s^2_{\rm R}=\frac{1}{T}\dif s^2_{\rm W}.
\end{equation}
Note that the above two kinds of  thermodynamic geometries are valid only for SBHs because they are based on 
the corresponding first law of thermodynamics. For RBHs, however, the first law of thermodynamics contains additional terms. Thus, we have to restrict the first law in a subphase space, and rewrite it, as an example, for a RBH described by a four-dimensional subphase space,
\begin{equation}
	\label{eq:wrong-1stlaw}
	\dif M=\frac{T}{1-\Pi}\dif S-\frac{P}{1-\Pi}\dif V, 
\end{equation}
where $\Pi$ is related to additional terms. 
In this case, if we constructed thermodynamic geometry from the problematic first law Eq. \eqref{eq:wrong-1stlaw}, the thermodynamic metric would also contain additional terms associated with $\Pi$ that has no thermodynamic counterpart. Correspondingly, for RBHs the conformal relationship between Ruppeiner geometry and Weinhold  geometry becomes
\begin{eqnarray}
	\dif s^2_{\rm R}=\frac{1-\Pi}{T}\dif s^2_{\rm W}.\label{metircR}
\end{eqnarray}

It is important to emphasize that the construction of thermodynamic geometry for RBHs is a challenging topic, and it must be based on the correct first law of thermodynamics. The microstructure of RBHs has also been studied~\cite{Tharanath:2014naa,NaveenaKumara:2020lgq,Pu:2019bxf} in different gravity theories with the help of thermodynamic geometry. As is known, RBHs are very special BH models and have been widely concerned~\cite{NaveenaKumara:2020jmk,Rizwan:2018ozh,Hennigar:2017apu,Nam:2019zyk} recently. Their first law of thermodynamics may need to be modified, and accordingly their thermodynamic geometry may be modified as well. It is hopeful to deeply understand the thermodynamics of RBHs from the perspective of thermodynamic geometry.

\subsubsection{Singularity of thermodynamic geometry and elimination}

A very special feature of charged AdS BHs is that the heat capacity at constant volume vanishes, i.e., $C_V=0$.  This property makes the thermodynamic line element, Eq.~\eqref{metircR}, singular and the corresponding thermodynamic information is unavailable from thermodynamic geometry. 
A new normalized scalar curvature is defined \cite{Wei:2019uqg},
\begin{eqnarray}
R_N=R\,C_V,
\end{eqnarray}
under which the divergence of Ruppeiner scalar curvature can subtly be eliminated.
For RBHs, extra terms will appear in thermodynamic line elements due to the modification in the first law of thermodynamics, which can be used to eliminate the divergence of thermodynamic scalar curvature. Therefore, RBHs can be viewed as an optional item to address the divergence of thermodynamic scalar curvature.

\section{Conclusion and outlook}
\label{sec:conclusion}

We have presented several topics of RBHs in the current review, 
where the importance of topics plays a major role in our choice. There are still three interesting issues we want to mention here. 
The first is whether  RBHs are trivial from the point of view of coordinate transformations, the second issue is what sources that generate rotating RBHs are, and the last issue is what differences between RBHs and SBHs are. 

For the first issue, it is known that every RBH corresponds to a metric with several coordinate singularities. 
Thus, one can always find a transformation to remove these singularities. 
For instance, the coordinate singularity of Hayward BHs \cite{Hu:2023iuw} can be removed in the Painlev\'e-Gullstand coordinate system \cite{Karl:2001rcs} and such a coordinate system is not unique, that is, the Kruskal–Szekeres coordinate and Eddington-Finkelstein coordinate systems  \cite{Wald:1984rg}  
have the same effect. We note that we just eliminate the singularity on 
an event horizon but not the event horizon itself.

The second issue arises due to the construction of rotating RBHs. 
Bambi and Modesto utilized \cite{Bambi:2013ufa} the NJA to create the metrics of rotating RBHs. 
On one hand, the NJA depends on theories of gravity. 
Taking a RBH model in the Chern-Simons gravity \cite{Moussa:2008sj} as an example, 
the NJA can provide a rotating metric, 
but the Pontryagin density, $^*RR$, does not vanish  \cite{Alexander:2009tp}, 
which means that the constructed metric will not satisfy equations of motion. 
On the other hand, as far as we know, 
there is still a lack of investigation on the classical field source of rotating RBHs.

At last, the most intriguing issue is possibly what the differences between RBHs and SBHs are. 
To answer this question, one must distinguish 
which differences are caused by the absence of singularities
and which are caused by the models’ characteristics, 
such as the differences between Schwarzschild and RN BHs.
This issue is expected to play a fundamental role in the study of RBHs.
Because the interior of a black hole cannot be directly observed, we have to investigate the dynamical and thermodynamic phenomena occurred outside horizons, in particular, the differences of phenomena with and without singularities.     
These differences may provide a chance to test RBHs from observational consequences mentioned in  Introduction, e.g., the gravitational wave, X-ray, and BH shadows, etc. 
The theoretical predictions together with currently available experimental data may shed some light on the existence of RBHs.

\section*{Acknowledgments}
This work is supported in part by the National Natural Science Foundation of China under Grant No.\ 12175108.

\appendix

\section{Segr\'e notation}
\label{app:segre}

Segr\'e notation (or Segr\'e symbol) is a systematic method for studying the complete intersection of two quadrics in algebraic geometry \cite{Dolgachev:2012cag}.
It is extended by Petrov \cite{Petrov:1954cg,Petrov:1961pej,Petrov:2000bs} to classify gravitational fields \cite{Stephani:2003tm}. 
However, the original and extended notations aim at different research subjects. The former targets second-order tensors in a vector space, 
and acts as an algebraic classification of Ricci tensors in general relativity, called the Segr\'e classification. 
The latter directs at fourth-order tensors in a bivector space
in general relativity, and acts as an algebraic classification of Weyl tensors, known as the Petrov classification.
The Segr\'e notation refers to the Segr\'e classification in our current context.
It provides a guide to searching the matter source in RBHs for a given metric and gravitational theory.

For two given tensors of order two, $\mathcal{R}$ and $\mathcal{I}$, which can also be regarded as two matrixes, 
one considers the $\lambda$-matrix, $\mathcal{R}-\lambda\mathcal{I}$, 
and computes the corresponding elementary divisors.
Supposing that $\det(\mathcal{R}-\lambda\mathcal{I})=0$ as an algebraic equation has $n$ distinct roots $\lambda_1,\ldots, \lambda_n$, we have the elementary divisors for every root $\lambda_i$,
\begin{equation}
(\lambda-\lambda_i)^{p_{i}^{(1)}}, \ldots, (\lambda-\lambda_i)^{p_{i}^{(s_i)}},\qquad
p_{i}^{(1)}\le\cdots\le p_{i}^{(s_i)},
\end{equation}
where $p_{i}^{(s_j)}$ is the multiplicity of eigenvalue $\lambda_i$ in the $s_j$-th divisor.
The Segr\'e notation is the collection,
\begin{equation}
\left[
(p_{1}^{(1)} \ldots p_{1}^{(s_1)} ) \ldots (p_{n}^{(1)} \ldots p_{n}^{(s_n)} )
\right].
\end{equation}

For a second-order EMT, one constructs an orthonormal basis $\hat{e}^\alpha_\mu$ \cite{Poisson:2009pwt} on the spacetime manifold $(\mathcal{M}, g_{\mu\nu})$, which satisfies  
\begin{equation}
g_{\alpha\beta} \hat{e}^\alpha_\mu \hat{e}^\beta_\nu= \eta_{\mu\nu},\qquad
\eta=\diag\{-1,1,1,1\},
\end{equation}
then decomposes the EMT by
\begin{equation}
\label{eq:emt-diagnal}
T^{\mu\nu} = \rho \hat{e}^\mu_0 \hat{e}^\nu_0
+p_1 \hat{e}^\mu_1 \hat{e}^\nu_1
+p_2 \hat{e}^\mu_2 \hat{e}^\nu_2
+p_3 \hat{e}^\mu_3 \hat{e}^\nu_3.
\end{equation}
In other words, the EMT in the basis $\hat{e}^\alpha_\mu$ can be represented as a diagonal matrix.
Thus, we can use the Segr\'e notation to classify the matter contents that have the possibility to generate non-rotating or rotating RBHs.

Due to the algebraic and physical properties of Eq.\ \eqref{eq:emt-diagnal}, the multiplicity is unit, 
i.e. we can use $1$ and the parentheses to denote every diagonal component of $T^{\mu\nu} $ 
and the equality among individual components. 
For instance, the algebraic property of EMT with $\rho=p_1$ and $p_2=p_3$ can be represented by $[(11)(11)]$. 
Here, we do not use commas to separate the time and space components.

\section{Zakhary-Mcintosh invariants}
\label{app:zm-invariants}


We list all the $17$ ZM invariants following Ref.\  \cite{Overduin:2020cif}. 

\begin{enumerate}
\item  Ricci data construed by Ricci tensors,
\begin{subequations}
\begin{equation}
    \ii_5=g^{\mu\nu}R_{\mu\nu},
\qquad
     \ii_6=R_{\mu\nu}R^{\mu\nu},
\end{equation}
\begin{equation}
     \ii_7\coloneqq R_{\mu}^{\;\;\nu}R_{\nu}^{\;\;\alpha}R_{\alpha}^{\;\;\mu},
\qquad
    \ii_8\coloneqq
    R_{\mu}^{\;\;\nu}R_{\nu}^{\;\;\alpha}R_{\alpha}^{\;\;\beta}R_{\beta}^{\;\;\mu},
\end{equation}
\end{subequations}
\item  Weyl data construed by Weyl tensors $W_{\alpha\beta_\mu\nu}$,
\begin{subequations}
\begin{equation}
    \ii_1\coloneqq W^{\mu\nu}_{\quad \alpha\beta}W^{\alpha\beta}_{\quad \mu\nu},
\qquad
    \ii_3\coloneqq W^{\mu\nu}_{\quad \alpha\beta}W^{\alpha\beta}_{\quad \rho\sigma}W^{\rho\sigma}_{\quad \mu\nu},
\end{equation}
\begin{equation}
    \ii_2\coloneqq -W^{\mu\nu}_{\quad \alpha\beta}W^{*\alpha\beta}_{\quad \mu\nu},
\qquad
    \ii_4\coloneqq -W^{\mu\nu}_{\quad \alpha\beta}W^{*\alpha\beta}_{\quad \rho\sigma}W^{\rho\sigma}_{\quad \mu\nu},
\end{equation}
\end{subequations}
\item  Mixed data constructed by both Ricci and Weyl tensors,
\begin{subequations}
\begin{equation}
    \ii_9
    \coloneqq W_{\mu\alpha\beta\nu}R^{\alpha\beta}R^{\nu\mu},\quad
    \ii_{10}
    \coloneqq -W^*_{\mu\alpha\beta\nu}R^{\alpha\beta}R^{\nu\mu},
\end{equation}
\begin{equation}
    \ii_{11}
    \coloneqq
    R^{\mu\nu} R^{\alpha\beta} \left(
    W_{\rho\mu\nu}^{\quad\; \sigma} W_{\sigma\alpha\beta}^{\quad\; \rho} 
    - W_{\rho\mu\nu}^{*\quad \sigma} W_{\sigma\alpha\beta}^{*\quad \rho}
    \right),
\end{equation}
\begin{equation}
    \ii_{12}
    \coloneqq
    -R^{\mu\nu} R^{\alpha\beta} \left(
    W_{\rho\mu\nu}^{*\quad\; \sigma} W_{\sigma\alpha\beta}^{\quad\; \rho} 
    + W_{\rho\mu\nu}^{\quad \sigma} W_{\sigma\alpha\beta}^{*\quad \rho}
    \right),
\end{equation}
\begin{equation}
    \ii_{13}
    \coloneqq 
    R^{\mu \rho} 
    R_{\rho}^{\;\; \alpha}
    R^{\nu \sigma} R_{\sigma}^{\;\beta}
    W_{\mu\nu\alpha\beta},\quad
     \ii_{14}
    \coloneqq 
    R^{\mu \rho} 
    R_{\rho}^{\;\; \alpha}
    R^{\nu \sigma} R_{\sigma}^{\;\beta}
    W^*_{\mu\nu\alpha\beta}
\end{equation}
\begin{equation}
    \ii_{15}
    \coloneqq
    \frac{1}{16} R^{\mu\nu}R^{\alpha\beta}
    \left(
    W_{\rho\mu\nu\sigma}W_{\;\;\alpha\beta}^{\rho\quad\sigma}
    +
    W^*_{\rho\mu\nu\sigma}W_{\quad\alpha\beta}^{*\rho\quad\sigma}
    \right),
\end{equation}
\begin{equation}
\begin{split}
    \ii_{16}
   & \coloneqq
   -\frac{1}{32} R^{\alpha\beta} R^{\mu\nu}
    \Big(
     W_{\rho\sigma\gamma\zeta}
    W^{\rho\quad\zeta}_{\;\;\alpha\beta}W^{\sigma\quad\zeta}_{\;\;\mu\nu}
    +W_{\rho\sigma\gamma\zeta}
    W^{*\rho\quad\zeta}_{\;\;\;\;\alpha\beta}W^{\sigma\quad\zeta}_{\;\;\;\;\mu\nu}\\
    &\qquad\qquad\qquad-W^*_{\rho\sigma\gamma\zeta}
    W^{*\rho\quad\zeta}_{\;\;\;\;\alpha\beta}W^{\sigma\quad\zeta}_{\;\;\mu\nu}
    +W^*_{\rho\sigma\gamma\zeta}
    W^{\rho\quad\zeta}_{\;\;\alpha\beta}W^{*\sigma\quad\zeta}_{\;\;\;\;\mu\nu}
    \Big),
\end{split}
\end{equation}
\begin{equation}
\begin{split}
    \ii_{17}
   & \coloneqq
   \frac{1}{32} R^{\alpha\beta} R^{\mu\nu}
    \Big(
     W^*_{\rho\sigma\gamma\zeta}
    W^{\rho\quad\zeta}_{\;\;\alpha\beta}W^{\sigma\quad\zeta}_{\;\;\mu\nu}
    +W^*_{\rho\sigma\gamma\zeta}
    W^{*\rho\quad\zeta}_{\;\;\;\;\alpha\beta}W^{*\sigma\quad\zeta}_{\;\;\;\;\mu\nu}\\
    &\qquad\qquad\qquad-W_{\rho\sigma\gamma\zeta}
    W^{*\rho\quad\zeta}_{\;\;\;\;\alpha\beta}W^{\sigma\quad\zeta}_{\;\;\mu\nu}
    +W_{\rho\sigma\gamma\zeta}
    W^{\rho\quad\zeta}_{\;\;\alpha\beta}W^{*\sigma\quad\zeta}_{\;\;\;\;\mu\nu}
    \Big),
\end{split}
\end{equation}
\end{subequations}
\end{enumerate}
where $W_{\mu\nu\alpha\beta}$ is Weyl tensor
and $W^*_{\mu\nu\alpha\beta}=\epsilon_{\mu\nu\rho\sigma}W^{\rho\sigma}_{\;\;\;\alpha\beta}/2$ denotes its dual.


\bibliographystyle{utphys}

\bibliography{references}

\end{document}